\documentclass{ws-ijmpa}

\newcommand{\ud}{\textrm{d}}
\newcommand{\uD}{\textrm{D}}
\newcommand{\mD}{\mathcal{D}}

\newcommand{\bigvev}[1]{{\left\langle #1 \right\rangle}}
 
\renewcommand{\bar}{\overline}
\newcommand{\tr}{\textrm{tr}}
\newcommand{\Tr}{\textrm{Tr}}
\newcommand{\gammap}{\gamma_5}

\newcommand{\beqa}{\begin{eqnarray}}
\newcommand{\eeqa}{\end{eqnarray}}
\newcommand{\beq}{\begin{equation}}
\newcommand{\eeq}{\end{equation}}
\newcommand{\vev}[1]{\left\langle\, #1\, \right\rangle}

\begin{document}

\markboth{Erich Poppitz and Yanwen Shang}
{Chiral lattice gauge theories via mirror-fermion decoupling: a mission {\tiny(im)}possible? }

%
%

\title{CHIRAL LATTICE GAUGE THEORIES VIA MIRROR-FERMION DECOUPLING: A MISSION
 { \footnotesize (IM)}POSSIBLE?}

\author{ERICH POPPITZ}

\address{Department of Physics, University of Toronto\\ 
 Toronto, ON, M5S1A7, Canada \\
poppitz@physics.utoronto.ca}

\author{YANWEN SHANG}

\address{ 
Perimeter Institute for Theoretical Physics \\ Waterloo, ON, N2J2W9, Canada\\
yshang@perimeterinstitute.ca}

\maketitle

\begin{abstract}
This is a review of  the status and outstanding issues in attempts to construct chiral lattice gauge theories by decoupling the mirror fermions from a vectorlike theory. In the  first half, we   
explain why studying nonperturbative chiral gauge dynamics may be of interest, enumerate the problems that a lattice formulation of chiral gauge theories must overcome, and briefly review our current knowledge.
We then discuss  the motivation and idea of mirror-fermion decoupling  and illustrate the desired features of the decoupling dynamics by a  simple solvable toy model. The role of exact   chiral symmetries and matching of 't Hooft anomalies on the lattice is also explained. The second,  more technical, half of the article is devoted to a discussion of the known and unknown features of mirror-decoupling dynamics formulated with Ginsparg-Wilson fermions. We end by pointing out possible directions for future studies.


\end{abstract}

\ccode{PACS numbers: 11.15.Ha, 11.15.Me, 11.30.Rd}

\bigskip

{{CONTENTS}}

\smallskip

{\footnotesize

1 Introduction \hfill 2

~~ 1.1 Chiral gauge theories through the ages \hfill 3

~~ 1.2 Where does the lattice stand? \hfill 4

~~ 1.3 Outline \hfill 5

2 Difficulties faced by chiral gauge theories on the lattice \hfill 7

~~ 2.1 An ambigiuous phase \hfill 7

~~ 2.2 The sign problem in chiral gauge theories \hfill 8

~~ 2.3 The fermion doubling problem \hfill 10

~~ 2.4 Ginsparg-Wilson fermions: an elegant solution\hfill 11

~~ 2.5 Problems with Ginsparg-Wilson fermions in chiral gauge theories\hfill 15

3 Chiral from vectorlike?\hfill 17

~~ 3.1 The idea of decoupling the mirror fermions\hfill 17

~~ 3.2 A toy model of decoupling in strong-coupling symmetric phases\hfill 19

~~ 3.3 Mirror global symmetries and 't Hooft anomaly matching\hfill 22

~~ 3.4 Comparison to ``pre-Ginsparg-Wilson" attempts of mirror/doubler decoupling\hfill 24

~~ 3.5 The big picture\hfill 26

~~ ~~ 3.5.1 Questions about the mirror dynamics\hfill  27

~~ ~~ 3.5.2 Answers---known and unknown\hfill 27

4 Theoretical and Monte Carlo studies of mirror decoupling via Ginsparg-Wilson fermions \hfill 28

~~ 4.1 Strong-coupling phases and properties of chiral partition functions\hfill 28

~~ ~~ 4.1.1 The ``1-0" model\hfill 29

~~ ~~ 4.1.2 Splitting the vectorlike partition function. The ``splitting theorem" \hfill 30

~~ ~~ 4.1.3 The phase structure of the ``1-0" model at strong mirror Yukawa coupling \hfill 32

~~ ~~ 4.1.4 The polarization operator and transversality \hfill 33

~~ ~~ 4.1.5 Further properties and exact relations obeyed by the mirror $\Pi_{\mu \nu}$ \hfill 35

~~ 4.2 Anomaly matching and its possible solutions \hfill 38

~~ 4.3 The symmetry arguments vs explicit lattice simulations \hfill 40

~~ 4.4 A different form of the split partition function; the relation to domain wall fermions \hfill 42

5 Outlook \hfill 47

}

\section{Introduction}	

To high energy theorists, a chiral gauge theory is not an unfamiliar concept. A Dirac spinor consists of two Weyl spinors of opposite chiralities: the left- and right-handed ones. A chiral gauge theory is a theory of   Weyl fermions, not necessarily forming  Dirac multiplets, but living in some complex representation of a Lie group and minimally coupled to the corresponding gauge field. We will adopt a broad definition of a ``chiral gauge theory"---a theory, where gauge-invariant mass terms for all 
fermions are not allowed---which includes  some interesting cases with pseudo-real representations, where fermion mass terms vanish due to statistics.

According to our experimentally verified knowledge, nature is described by just such a theory. It might sound a bit mysterious   that we do not yet have a  method of approximating an arbitrary chiral gauge theory by latticizing and then simulating it on a computer---even in principle.  If one wishes to use the lattice to gain some insight  into its non-perturbative properties, we just do not know how to do it.

Apart from interest in physics of  the Standard Model---which, at low energies, is a weakly-coupled spontaneously broken chiral gauge theory that does not 
obviously call for a lattice study---interest in strong chiral gauge dynamics  has both intensified and abated during the past few decades. From the overview in the next  Section,  it should be clear that  while there  exist potential applications of strong chiral gauge dynamics to particle physics, at the moment it appears difficult to identify ``the" chiral  theory   most  relevant to particle physics model-building (apart from the weakly-coupled Standard Model, of course). Thus,  the problem of a lattice formulation of chiral gauge theories is currently largely of theoretical interest. This may or may not change after the LHC data is understood. Regardless, we find the problem sufficiently intriguing\footnote{Leading experts' published opinions on this matter vary, e.g.:
``If a solution to putting chiral gauge theories on the lattice proves to be a complicated 
and not especially enlightening enterprise, then it probably is not worth the effort 
(unless the LHC finds evidence for a strongly coupled chiral gauge theory!)." (D.B. Kaplan)\cite{Kaplan:2009yg}, or: 
``Without a proper 
lattice formulation of a chiral gauge theory, it is unclear whether such models 
make any sense as fundamental field theories." (M. Creutz)\cite{Creutz:2004eq}.
 } to devote some effort to its study.

\subsection{Chiral gauge theories through the ages}
\label{throughtheages}

Strongly  coupled chiral gauge theories were first studied in the 1980's for their possible relevance to  quark  and lepton compositeness. To illustrate both the connection to such ideas and the quite different physics that can occur in chiral gauge theories (compared to  vectorlike, or ``QCD-like" theories),  
consider  the classic example\cite{Dimopoulos:1980hn} of a four-dimensional strongly-coupled chiral gauge theory. This is an $SU(5)$ gauge theory with a ${\bf 5^*}$ and a ${\bf 10}$ Weyl-fermion representations, both taken left-handed. The  theory has one anomaly-free chiral $U(1)$ global symmetry,\footnote{The quadratic Casimir is $1/2$ for the fundamental  and $3/2$ for the antisymmetric tensor of $SU(5)$.} under which the ${\bf 5^*}$  fermion has charge $-3$ and the ${\bf 10}$ has charge $1$. It is a simple matter to check that the 't Hooft anomaly matching\cite{'tHooft:1980,Frishman:1980dq,Coleman:1982yg}  conditions
 for  the $U(1)^3$ and $U(1)$ anomalies are saturated by a single massless composite Weyl fermion, a $\bf 5^*$-$\bf 5^*$-$\bf 10$ bound state. Thus, in the absence of  gauge-invariant order parameters to break the chiral symmetry, the  likely  infrared (IR)  dynamics is that of a free   single massless gauge-singlet composite fermion (a ``tumbling" scenario based on ``most-attractive-channel" (MAC) reasoning is  consistent with such an  IR spectrum\cite{Raby:1979my}). 
Since the anomaly-free chiral symmetry is unbroken,  this chiral gauge theory  exhibits  ``confinement without chiral symmetry  breaking"---clearly, an IR dynamics   different  from that of QCD-like theories (at least from vectorlike theories without supersymmetry and the associated massless elementary 
scalars: it turns out that supersymmetric QCD, a vectorlike theory, has phases that exhibit confinement without chiral symmetry breaking, see\cite{Strassler:2001px} for an overview and references). 
 
 The appearance of massless composite fermions in many  chiral gauge theories, as suggested by anomaly matching  and the (non rigorous) MAC arguments gave rise to many attempts to build ``preon" models for elementary particle physics; see\cite{Rosner:1998wh} for  references. This direction is currently not being actively pursued (at least not via direct studies of four-dimensional strong gauge dynamics, see below), not because compositeness is ruled out\footnote{Experimental bounds are stronger than in the 1980's, constraining the scale of compositeness\cite{Amsler:2008zzb} to be in the multi-TeV region.}, but because constructing realistic models turned out to be difficult and 
 because the theoretical tools available, at least for nonsupersymmetric models, are limited to the already mentioned 't Hooft consistency conditions and MAC. 
  However, the idea that some of the particles we see may be composite is still alive.   More recently,  after the advent of the AdS/CFT correspondence\cite{Aharony:1999ti},  models of electroweak-scale physics have used   quarks, leptons, and gauge fields ``living" at various locations in slices of AdS$_5$. These phenomenological models are thought to be holographically dual to strongly-coupled gauge theories, where some of the Standard Model particles arise as (partially) composite objects\cite{Gherghetta:2006ha}.  However, it is currently not known what are the four-dimensional gauge theories whose dual description  is given by these phenomenological warped models, much less whether they are chiral or not.

 Another surge of interest in chiral gauge dynamics occurred in the supersymmetric context in the mid-1990's. It was due to the then-held belief that only chiral gauge theories can be used to dynamically break supersymmetry. Recall that if supersymmetry is relevant to particle physics, it has to be broken---hence, the mechanism of supersymmetry breaking is of interest to superparticle phenomenology.
A number of  chiral supersymmetry-breaking theories were studied, see\cite{Poppitz:1998vd} for a review,  using the newly available theoretical tools of holomorphy and duality. The interest in  chiral supersymmetric theories has somewhat subsided after the recent discovery, reviewed in\cite{Intriligator:2007cp}, that vectorlike supersymmetric theories can have metastable supersymmetry-breaking vacua, sufficiently long-lived  for particle physics model-building purposes. 

Electroweak symmetry breaking and fermion mass generation is another area where strong chiral gauge dynamics has been invoked\cite{Terning:1994sc,Holdom:1996mt}, for example using\cite{Appelquist:2003uu}    tumbling (via MAC) in chiral gauge theories to construct extended technicolor models. A more recent rather speculative 
proposal\cite{Csaki:2010cs} is to break electroweak symmetry in a chiral theory of massless fermions (a fourth generation) with both electric and magnetic $U(1)$ charges. Consistent theories of ``mutually nonlocal" massless monopoles and dyons have been argued to exist in ${\cal{N}}$$=$$2$ supersymmetry\cite{Argyres:1995jj},  but it is not known whether nonsupersymmetric versions exist; addressing this  on the lattice is challenging, as such theories are inherently strongly coupled. It is difficult to assess  the viability and study  detailed features of the proposed strong chiral electroweak-breaking models, without credible  tools to study their dynamics\footnote{Despite the new poorly-understood strong chiral  interactions at the $TeV$-scale, characteristic\cite{Holdom:2010za,Csaki:2010cs}  signatures of these models could be found at the LHC or their absence used to rule them out.}.

Finally,  on the theoretical side, there has been little progress (in the nonsupersymmetric context) in the nonperturbative understanding of chiral gauge dynamics since the 1980's\cite{Eichten:1985fs}. An exception is  the recent work\cite{Shifman:2008cx,Poppitz:2009kz,Poppitz:2009uq} on studying the nonperturbative dynamics of chiral gauge theories  compactified on $R^3 \times S^1$. At   sufficiently small radius of $S^1$, semiclassical methods can be used  to study the generation of mass gap in the gauge sector (i.e., the string tensions) and the breaking of discrete or abelian chiral symmetries. These studies show, once again,  that nonsupersymmetric chiral gauge dynamics is different from that of QCD-like theories. However, while instructive, the validity of the semiclassical approach does not   extend   to the physical  case of  decompactified $R^4$.
 
 \subsection{Where does the lattice stand?}
 
  Ideally, lattice studies should provide both a definition and a  means to study the dynamics of any strongly-coupled gauge theory. In vectorlike nonsupersymmetric theories, significant theoretical and practical progress has been made in the past 15 years. In chiral theories, there are both conceptual and technical obstacles. These, along with some of the  ongoing attempts to overcome them, are the subject of this review. 
 
 We note that 
there has been some recent progress, reviewed in\cite{Catterall:2009it}, towards a lattice definition  and even numerical studies  of some of the  theories of possible phenomenological interest mentioned in  Section \ref{throughtheages}, most notably the ones with supersymmetry. Practical progress is, however,  limited mostly to vectorlike supersymmetric theories with extended supersymmetry  in lower dimensions; 
see\cite{Giedt:2009yd} for an exception. The chiral supersymmetric case, in both two (2d) and four (4d) dimensions is currently out of reach. As already stated above,
for nonsupersymmetric chiral theories, the lack of  theoretical tools to study such theories makes it difficult to assess the various  models. 
  
 The long-standing issue of chiral lattice gauge theories has been previously reviewed in\cite{Golterman:2000hr,Luscher:2000hn,Neuberger:2001nb}, where different approaches    and their status are discussed in more detail than we can go into here. We also recommend the already quoted 
lectures\cite{Kaplan:2009yg}  on exact chiral symmetry  in vectorlike lattice theories; we will not review this subject in great detail and will largely assume  the reader's familiarity with it.

\subsection{Outline}

\begin{enumerate}
\item
In Section~\ref{difficulties}, we describe the problems that must be overcome in order to define chiral symmetry and chiral gauge theories on the lattice.
\begin{itemize}
\item In Section~\ref{ambiguousphase}, we recall why the
partition function of a chiral gauge theory is only defined up to a phase, a property not usually relevant in the continuum, but important for the lattice definition. 
\item In 
Section~\ref{signproblem}, we briefly discuss the complexity of chiral fermion determinants (one of the main differences between chiral and vectorlike theories from the point of view of the Euclidean path integral).
\item In Sections~\ref{doublingproblem} and~\ref{GWsolution}, we  recall the fermion doubling problem, its solution in terms of ``Ginsparg-Wilson fermions," and  some of their pertinent features.  
\item In Section~\ref{GWproblems}, we list the various problems arising when using Ginsparg-Wilson fermions to define chiral lattice gauge theories.
\end{itemize}
\item
In Section~\ref{chiralfromvectorlike}, we give a non-technical discussion of the  ``mirror-decoupling" approach to chiral lattice gauge theories. 
\begin{itemize}
\item In Section~\ref{ideaofdecoupling},  the main idea is described using continuum notation, along with a discussion of some motivational evidence. 
\item In Section~\ref{toymodel}, we explain the nature of the strong-coupling symmetric phases in lattice  theories with strong multi-fermion or Yukawa interactions and illustrate the desired features of the mirror-decoupling dynamics with the help of  a simple solvable toy model. 
\item In Section~\ref{mirrorsymmetries}, we use 't Hooft anomaly matching arguments to explain why breaking all mirror chiral global symmetries, anomalous or  not, is a necessary condition for mirror-fermion decoupling.
\item 
In Section~\ref{preginspargwilsoncomparison}, we compare our approach to ``pre-Ginsparg-Wilson" attempts of mirror/doubler decoupling. We emphasize  that the crucial difference lies in the absence of  exact lattice chiral symmetries in earlier implementations. 
\item In Section~\ref{generaldiscussion}, devoted to the ``big picture," we end the more descriptive part of the review by giving  a list of questions about  the mirror-decoupling dynamics that need to be answered and  a summary   the (known and unknown) answers. 
\end{itemize}
\item In Section~\ref{implementation}, we  study  the many technical issues that arise in the Ginsparg-Wilson-fermion implementation of  ``mirror-decoupling" approach.
\begin{itemize}
\item In Section~\ref{chiralZ}, we study general properties of chiral partition functions. Many theoretical results, more generally valid, will be illustrated on the example of a simple model, as discussed below. The tools developed allow the  questions listed in Section \ref{generaldiscussion} to be addressed,  here and in future work.
\begin{itemize}
\item In Section~\ref{10model}, we  introduce the model that we have   studied extensively, the ``1-0" 2d model.  Here, the ``mirror" fermions are in an anomalous representation, so we do not expect to decouple them without breaking the gauge symmetry. However, this model provides a useful playground to study the dynamics of strong Yukawa  or multifermion interactions with Ginsparg-Wilson fermions via inexpensive numerical simulations and allows  addressing some of the  questions of Section~\ref{generaldiscussion}.
\item In Section~\ref{splitZ}, we  explain how to separate the action and measure of a vectorlike theory into   ``light" and ``mirror" parts and discuss the ``splitting theorem" about the dependence of chiral partition functions  with general interactions  on the gauge background. This is an important analytical result, without   which   the studies discussed here would have not been possible.   
\item In Section~\ref{phasestructure} we use the split partition function to study the phase structure of the mirror dynamics of the ``1-0" model at strong Yukawa coupling.
\item In Sections~\ref{transversality10} and~\ref{mirrorpimunu}, 
we study the mirror polarization operator in vanishing gauge background. Its real part is a universal probe of the mirror-fermion spectrum, while its imaginary part contains (in 2d) the mirror-fermion anomaly. 
We list 
  some general  exact  (i.e., independent of the nature and strength of the mirror interactions) properties of the mirror polarization operator, useful for its analytic and numeric studies.
    \end{itemize}
 \item
In Section~\ref{matching} we
 explain how to formulate and study 't Hooft anomaly matching on the lattice, applied to the case of (strong or weak) mirror dynamics.
  \item
In Section~\ref{symmetryvsexplicit}, we discuss the results of the numerical study of the realization of anomaly matching in the  mirror sector of the ``1-0" model, both in the  strongly-coupled symmetric and weakly-coupled  ``broken"  phase. We show that in the strong symmetric phase  anomaly matching is satisfied with the minimal number of massless charged fermions (and stress the role of the Majorana coupling in ensuring minimality).
\item
In Section~\ref{domainwall}, we rewrite the partition function in a different basis, motivated by the relation of Ginsparg-Wilson to domain-wall fermions. This representation is useful to further understand aspects of the dynamics: we  explain the appearance of extra massless vectorlike fermions, found in Section \ref{symmetryvsexplicit}, when one of the mirror-theory couplings (the Majorana coupling) vanishes. We also  comment on   the relation of our formulation to other  proposals  to use strong coupling in the domain wall set-up to decouple mirror fermions.
  \end{itemize}
\item
In Section~\ref{summary}, we conclude with a brief  discussion of the outlook for the future. 
\end{enumerate}

\section{Difficulties faced by chiral gauge theories on the lattice}
   \label{difficulties}
   
 \subsection{An ambiguous phase}
 \label{ambiguousphase}
 
Before we explain why defining lattice chiral gauge theories is so hard, we should first mention that the partition function of any chiral theory is, rigorously speaking, not defined. This is because an operator that maps between two independent vector spaces does not have a natural definition of its determinant.  Take for example the kinetic term of a pair of left-handed Weyl fermions:
 \begin{equation}
 \label{s1}
S=\int d^4 x \bar{\psi}_{L}D\psi_{L}
 \end{equation}
where $D$ stands for the usual Weyl operator and $\bar\psi_{L}$ and $\psi_L$ are two independent left-handed Weyl fermions (in Euclidean space, the left-handed spinors and their conjugates transform in different representations of the Euclidean ``Lorentz" group $SO(4) \simeq SU(2) \times SU(2)$, not related by complex conjugation as in Minkowski space). Attempting to define the Grassmann path integral:
 \begin{equation}
 \label{s2}
Z=\int D \bar{\psi}_{L} D\psi_{L} e^{S},
 \end{equation}
one imagines choosing a set of orthonormal basis vectors, say ${u_i}$ and ${v_i}$, and expanding in them the fermion fields with Grassmann coefficients ($\bar\psi_L$ is expanded in $u^\dagger_i$ and $\psi_L$---in $v_i$, see Section \ref{splitZ}). The subscript $i$ runs from 1 to the dimension of the fermion phase space, say  $d$, which we for simplicity assume to be the same for    $\bar\psi_{L}$ and $\psi_{L}$. Certainly $d$ is infinite for theories defined in the continuum spacetime, and this discussion is rather formal at the moment. The partition function (\ref{s2}) is evaluated by just defining $Z=\det (u^\dag_i D v_j)$, where $u^\dag_i D v_j$ is a $d\times d$ matrix (of indices $i$ and $j$), whose determinant is always defined.  
Notice, however, that the bases $\{u_i\}$ and $\{v_i\}$ are not unique. We can equally well choose, instead of, e.g., ${v_i} $, a different set of vectors $v_i'=U_{ij} v_j$, where $U_{ij}$ is a unitary matrix. If we compute $Z$ as above but use the new vectors ${v'_i}$, we find the result differs by a factor of $\det U_{ij}$. This factor is always a pure complex phase since $U_{ij}$ is unitary. The argument presented here is  equivalent to the trivial fact that the identification between two vector spaces of the same dimension is not unique.

Such a phase ambiguity of the partition function is usually not a severe problem since it is sufficient to just choose and then stick to a particular set of basis vectors. The ambiguous phase of $Z$---provided it can be chosen independently of the gauge field, as in perturbative continuum regularizations---is always divided out in the vacuum expectation values of any operator, which are the  physical observables. 
In what follows, we will see that in attempts to define chiral lattice gauge theories  this ambiguous phase can not be taken to be gauge-field independent. Its determination constitutes the so-called ``measure problem," that will play an important role in our further discussion.

\subsection{The sign problem in chiral gauge theories}
\label{signproblem}

Even if the ``measure problem" of Section \ref{ambiguousphase} is solved, an important issue  that remains is the sign problem of chiral partition functions.
The sign problem is likely to hinder an immediate application of any successful lattice definition of chiral gauge theories to studies of strong gauge dynamics via Monte Carlo methods (as it   generically occurs in 4d chiral gauge theories). 
However, as we discuss at the end of this Section, there are some exceptions---the continuum fermion determinant is real in some special cases in 4d  as well as in 2d anomaly-free chiral gauge theories.

  From the point of view of the Euclidean path integral, generic chiral and vectorlike theories  differ  in  that in the vectorlike case the fermion effective action is real, while  in the generic chiral case it has an imaginary part. The real part of the fermion effective action in a chiral gauge theory  is  equal  to one-half that of the vector theory obtained by adding massless mirror fermions to the chiral theory. 
 The appearance of an imaginary part in chiral theories can be explained in the continuum and we will briefly do so in this Section.  In the Euclidean path integral language, the imaginary part\footnote{To avoid confusion, we note that after integration over the gauge fields, the partition function is real---the imaginary contributions to the partition function of field configurations that are parity images of each other cancel (while the real parts are equal) leaving only a sign problem\cite{Hsu:1995qm}.} is the one that must be responsible, along with the distinct effects\cite{Shifman:2008cx,Poppitz:2009uq}    of nontrivial topological sectors, for the quite different dynamics of strongly-coupled vectorlike and chiral gauge theories (as seen in our discussion of the $SU(5)$ chiral model in Section \ref{throughtheages}). 
 
The reason for the sign problem is that chiral gauge theories---such as the Standard Model---are left-right asymmetric, i.e. violate parity. In Euclidean space, the parity-violating part of the effective action for fermions is purely imaginary. One can see this in perturbation theory by recalling that   the fermion determinant can be represented by a sum of one-loop fermion graphs with an arbitrary number of  gauge field insertions.
The real, parity-even, part of the effective action sums over  loops of vectorlike fermions with  the quantum numbers of the left- and right-handed fermions of the chiral gauge theory:\footnote{To  obtain   the formal expressions  (\ref{realpart}) and (\ref{impart}), one expands log$Z[A]_\pm$$=$${\rm log\,det } D_\pm$ (with $D_\pm$$=$$D_0 - i \gamma_\mu A_\mu {1 \pm \gamma_5 \over 2}$, $D_0 = \gamma^\mu \partial_\mu$) in a perturbative series and separates the real and imaginary parts; we use hermitean Euclidean gamma matrices. We note also that similar expansions for the real and imaginary parts can be obtained  using the lattice  definition of the determinant via the Neuberger-Dirac operator; naturally, these coincide with eqns.~(\ref{realpart}), (\ref{impart}) if the naive continuum limit is taken first.}  
\begin{equation}
\label{realpart}
{\rm Re\; log \; Z[A]_\pm} = -  {1 \over 2}  \sum\limits_{n=1}^\infty {1 \over 2 n} \; {\rm Tr} \; (D_0^{-1} i A^\mu \gamma_\mu)^{2n}~+ {\rm c.t.},
\end{equation}
where $D_0$ is the free Dirac-fermion massless propagator  (we give   the effective action for  a single Weyl fermion;  the expression (\ref{realpart})  has to be summed over all Weyl fermions of the theory).  
The imaginary part of the effective action for a left- or right-handed Weyl fermion, on the other hand, is given by the parity-odd expression: \begin{equation}
\label{impart}
i\; {\rm Im \; log }\; Z[A]_\pm = \mp {1\over 2}   \sum\limits_{n=0}^\infty {1 \over 2 n+1} \;  {\rm Tr} \;\gamma_5(D_0^{-1}  i A^\mu \gamma_\mu)^{2n+1} + {\rm c.t.}~.
\end{equation} 
Both  (\ref{realpart}) and (\ref{impart}) are valid up to the addition of local counterterms,  denoted by ``c.t.," which ensure that the effective action is finite and gauge invariant (if gauge anomalies cancel). 

In 4d chiral gauge theories, the $n>1$ terms in the imaginary part of the effective action are finite and gauge invariant---and must contain important information about the chiral gauge theory dynamics, as we stipulated above. 
Using Pauli-Villars or zeta-function regularization, one can show\cite{AlvarezGaume:1985di} that the difference between the imaginary part of the fermion effective action (\ref{impart}) computed for two different perturbative gauge backgrounds,   $A$ and $A_0$, in the anomaly-free case, can be written as:
\begin{equation}\label{eta}
  {\rm Im \; log } Z[A] -    {\rm Im \; log }  Z[A_0]  =   \pi \eta[0]~.
\end{equation} 
Here, $\eta[0]$ is  the ``$\eta$-invariant," or spectral asymmetry, of a  five-dimensional hermitean Dirac operator $\hat H$ constructed from the the four-dimensional vectorlike Dirac operator ($A_\mu = A_\mu^a T^a$ is hermitian): 
\begin{equation}
\label{5doperator}
\hat{H} = \gamma_5 i {\partial \over \partial t} + i D[A(t)] , ~ D[A] =D_0 - i \gamma_\mu A_\mu~,
\end{equation} where $A(t)$ smoothly interpolates between $A(-\infty) = A_0$ and $A(\infty) = A$. The $\eta$-invariant $\eta[0]$ appearing in (\ref{eta}) is  the analytic continuation  to $s=0$ of $\eta[s]$, defined   for  large Re$(s) > 0$ as:   
 \begin{equation}
 \label{etas}
 \eta[s] \equiv  {\rm Tr}\; \frac{\hat{H}}{   (\hat{H}^2)^{{1\over 2}+s} }~.\end{equation}
The definition has to be slightly modified when zero modes of $\hat{H}$ are present.\cite{AlvarezGaume:1985di}  

 The $\eta$-invariant representation of the imaginary part of the chiral determinant is  useful to study anomaly cancellation, as it can be shown that the change of the effective action (\ref{eta}) under gauge transformations of $A$ is a topological invariant\cite{Witten:1985xe,DellaPietra:1986qd}.  On the other hand, the imaginary part of the action itself is not a topological term   and relating it to the $\eta$-invariant, computed with some fiducial background field, e.g., $A_0$ in (\ref{eta}),  sheds  no light on  dynamical aspects of the physics.

Another use of equations (\ref{eta}) and (\ref{etas}) is to quickly argue that the imaginary part of log$Z[A]$ vanishes for real representations.
  This  follows from the fact that for real representations, there exists a gauge group element $\sigma$ such that $(i A_\mu)^* =  \sigma^{-1} i A_\mu \sigma$. If $\Psi$ is an 
eigenfunction of $\hat{H}$, eqn.~(\ref{5doperator}), with eigenvalue $\lambda$, then $\Psi^\prime = 
B^{-1} \sigma^{-1} \Psi^*$, with $B^{-1} \gamma_\mu B =  \gamma_\mu^*$, has eigenvalue\footnote{This follows from  $\sigma B  \hat{H} B^{-1} \sigma^{-1} = - \hat{H}^*$, which is easy to check.}  $- \lambda$. Consequently $\eta[s]$, eqn.~(\ref{etas}), vanishes for any $s$, and so does  the spectral asymmetry $\eta[0]$ and the imaginary part of the fermion effective action.

 The result of the previous paragraph is of interest for some chiral gauge theories. There exists a 4d  chiral  theory that has a real positive determinant---the chiral $SU(2)$ gauge theory with a single Weyl fermion in the three-index symmetric  representation. Despite 
the pseudo-reality of the $j=3/2$ representation, a fermion mass term vanishes by Fermi statistics and the theory is usually referred to as chiral.\footnote{The  supersymmetric version of this theory has been used as a model for dynamical supersymmetry breaking, but recent arguments\cite{Intriligator:2005if,Poppitz:2009kz} suggest that it is conformal in the IR; the nonsupersymmetric version is believed to confine and break its discrete anomaly-free chiral 
symmetry.\cite{Poppitz:2009kz} }

In 2d anomaly-free chiral models, the terms  with $n=0$ in the imaginary part (\ref{impart}) cancel, while terms with  more than two powers of the gauge field in the Weyl determinant (the sum of (\ref{realpart}) and (\ref{impart})) can be argued to vanish\cite{Jackiw:1984zi} due to their  regulator-independence, gauge invariance, and chirality. Thus, at least in the continuum, 2d anomaly-free chiral gauge theories can be defined to have a real determinant. That this is so becomes explicit  upon examining their solution via bosonization\cite{Halliday:1985tg,Kutasov:1994xq}. 

We expect that the positivity of the fermion determinant should  be reproduced  in the continuum limit by a  lattice definition, hence (assuming  that this implies a less severe sign problem on the lattice) a successful proposal for defining chiral lattice gauge theories stands a chance to be tested  via Monte Carlo simulations  with dynamical gauge fields only in 2d or in some special 4d  chiral gauge theories, like the $j=3/2$ $SU(2)$ theory mentioned above---at least in the foreseeable future.

\subsection{The fermion doubling problem}
\label{doublingproblem}

As a most simple-minded attempt to define lattice theories of fermions, one may just discretize the Dirac or Weyl operator as   one would normally do for the Laplacian. Not surprisingly, this does not work, or we would not have to write this review. The problem is that when one takes the continuum limit of   a lattice theory with a naive discretization of the Dirac or Weyl operator, one finds, most intriguingly, that every Weyl fermion introduced in the action is accompanied by an unexpected massless doubler of opposite chirality.

One can most easily see that this  must happen by recalling the axial anomaly.  As is well known, in quantum field theory (QFT) with Dirac spinors, even if the action is invariant under the axial  rotation:
\begin{equation}
\label{chi1}
\bar\psi\rightarrow\bar\psi e^{i\alpha \gamma_5}\qquad \psi\rightarrow e^{i\alpha \gamma_5}\psi,
\end{equation}
 the partition function is, in general, not invariant. One way to understand this is due to 
 Fujikawa\cite{Fujikawa:1980eg}, 
  who pointed out that although naively such a field rotation should certainly leave the  fermion measure unchanged (as one expects that  $\rm{tr}\; \alpha(x)\gamma_5=0$ for arbitrary $\alpha(x)$), because the fermion phase space is infinitely dimensional, $d=\infty$ in our notation, it needs to be regularized and there exists no method of doing so respecting both the vector and the axial symmetry simultaneously. 
 
 Trying to repeat the same argument on the lattice immediately leads to problems. On a finite lattice, the theory is perfectly regularized and $d$ is a finite integer. We apparently find a regularization scheme of the measure, which is invariant under both vector and axial rotations, leaving absolutely no room for an anomaly to arise. Applying this logic to gauge symmetries and gauge anomalies in chiral theories, we would similarly conclude that  gauge anomalies do not exist on the lattice. Since in the continuum left- and right-handed Weyl fermions contribute to the anomaly with opposite signs and  cancel if their charges are equal, the absence of anomalies on the lattice is only possible if every Weyl fermion that survives in the continuum limit is accompanied by a doubler  with precisely the same charge but of opposite chirality. In other words, the fermion species are automatically doubled.  The impossibility of preserving an exact chiral symmetry of the form (\ref{chi1}) on the lattice without fermion doubling is the (in)famous fermion doubling problem. 

\subsection{Ginsparg-Wilson fermions: an elegant solution}
\label{GWsolution}

According to the above discussion, it should be clear that the only way to avoid the fermion doubling problem on the lattice is to explicitly break the axial (or chiral) symmetry slightly so that the action is only approximately invariant. One hopes that  in the continuum limit, this approximate symmetry can be fine-tuned to be exact and that the anomaly is properly reproduced as a remnant of the explicit breaking at the cutoff scale, see\cite{Karsten:1980wd}.
 Methods of this kind exist and are widely used in vector-like theories, e.g. QCD.

However, having only an approximate symmetry in lattice QFT  has many disadvantages. For our purposes, the most important is that it forbids pure chiral actions---or any action with a vectorlike fermion content (meaning equal number of left- and right-handed Weyl fermions with the same charge so they form vector multiplets), but consisting of two separated chiral sectors. Because  the  chiral symmetry is explicitly broken by the regulator, fermions with opposite chiralities must necessarily couple.

Ginsparg and Wilson (GW) proposed\cite{Ginsparg:1981bj} an elegant alternative in 1982. They suggested replacing the Dirac operator on the lattice by the so called GW operator, which satisfies the following two conditions:
\begin{equation}
\label{GW1}
\{D, \gamma_5\}=a D\gamma_5 D, \quad (\gamma_5 D)^\dag=\gamma_5 D.
\end{equation}
Here, $a$ is the lattice spacing. In the continuum limit $ a\rightarrow 0$, the rhs of the above anti-commutator vanishes and $D$ approaches the usual Dirac operator.

Using the Ginsparg-Wilson operator in the fermion kinetic terms is sufficient to eliminate the fermion doubling problem. It works, of course, because $D$ explicitly breaks the chiral symmetry (as it does not anticommute with $\gamma_5$). What is really beautiful, however, is that the GW operator allows to define an exact chiral symmetry on the lattice. To see this, define a Ònew $\gamma_5$Ó matrix by:
\begin{equation}
\label{GW2}\hat\gamma_5=(1-aD)\gamma_5,\end{equation}
and observe that (\ref{GW1}) implies that the following two equations are exact:
\begin{equation}
\label{GW3}
\hat\gamma_5 D+D\gamma_5=0, \quad \hat\gamma_5^2=1.
\end{equation}
The first equation says that actions like $S=\bar\psi D\psi$ are invariant under the ÒGW axial rotationÓ:
\begin{equation}
\label{GW4}
 \bar\psi\rightarrow \bar\psi e^{i\alpha\hat\gamma_5},\quad \psi\rightarrow e^{i\alpha\gamma_5}\psi,
\end{equation}
which is an exact symmetry of the action at a finite lattice spacing. In the continuum limit when the lattice spacing vanishes, $\hat\gamma_5\rightarrow\gamma_5$ and the traditional axial rotation (\ref{chi1}) is recovered.

We   note that while the GW relation was proposed in 1982, no explicit form of an operator obeying (\ref{GW1}) was known until 1997, when it was found by Neuberger as a result of a remarkable development\cite{Kaplan:1992bt,Kaplan:1992sg,NN1,NN2,Neuberger:1997fp}. We will not need the explicit form of the GW operator in this review, but only the properties (\ref{GW1}-\ref{GW4}) it satisfies;  see\cite{Kaplan:2009yg} for a recent introduction.

Now,   $\hat\gamma_5^2=1$ implies that $\frac{1\pm\hat\gamma_5}{2}$ define two projection operators, and therefore one can build, out of Dirac spinors, the ``GW chiral fermions" as  follows:\begin{equation}
\label{GW41}
\bar\psi_{L/R}\equiv\bar\psi\frac{1\mp\hat\gamma_5}{2},\quad \psi_{L/R}\equiv \frac{1\pm \gamma_5}{2}\psi.\end{equation}
This definition for the spinor $\psi_{L/R}$ is identical to the usual one in the continuum, but for $\bar\psi$, it is slightly different---but the difference disappears in the continuum limit. 
Thus, one can define ``chiral" theories on a finite lattice, free of fermion doubling, using these ÒGW chiral fermionsÓ. For example, a single massless free fermion can be described by the action $S=\bar\psi_L D\psi_L$, where $\bar\psi_L$, defined by (\ref{GW41}), is not the usual Weyl fermion that we are familiar with in the continuum. If there are no gauge fields, the measure can also be easily defined, as we will discuss later. The difficulties that arise when gauge fields are present are discussed   in Section \ref{GWproblems} below. 

Let us now explain how the correct chiral anomaly is derived in the GW formalism.  
 In a sense, the anomaly works out just as Fujikawa pointed out in the continuum: the anomaly exists because the fermion measure is not invariant. 
To see how the index theorem works out in vectorlike
theories, let us again start with the partition function for a single Dirac fermion:
\begin{equation}
\label{eq:dirac_action}
\mathcal{Z}=\int\mD\bar\psi\mD\psi \exp\left(\bar\psi\uD\psi\right)\,,
\end{equation}
where $\psi$ and $\bar\psi$ are both Dirac spinors.  In any vectorlike theory,
the fermion measure in the path integral is trivially defined on lattice. 
One simply chooses any orthonormal basis of convenience 
(momentum eigenstates or lattice-site eigenstates, for example), expands
the spinors with the chosen basis vectors with Grassmann expansion
coefficients, and defines the fermion integral
  by the usual Grassmann integral.
When the lattice has a finite size, such an orthonormal basis is always 
finite and even dimensional, with dimension we denote as $d$.  
The operator $\uD$ in \eqref{eq:dirac_action}
is the GW operator introduced before, satisfying the conditions \eqref{GW1}.
From now on, we implicitly understand that some  background gauge field
is present and the GW operator $\uD$ is a functional of
the gauge field configuration, because it must approach  the covariant
Dirac operator in the continuum limit. As already mentioned, the explicit form of $\uD$ can be
complicated  and is not needed for the general discussion here.
We also set the lattice spacing $a=1$ from now on.

In the four-dimensional Weyl representation of $\gamma$-matrices, $\gammap$ is given by:
\begin{equation}
\label{g5}
(\gammap)_{xx}=\left(
\begin{array}{cc}\mathbf 1&\\&-\mathbf 1\end{array}\right),
\qquad x\textrm{ is any lattice site}, 
\end{equation}
and $(\gammap)_{xy}=0$ if $x\ne y$. On the other hand,  $\hat\gammap$ 
is defined through equation
\eqref{GW2}. Again, it suffices for now to recall that
$\hat\gammap^\dag=\hat\gammap$ and $\hat\gammap^2=1$ (following from
the GW relations) and that $\hat\gammap\rightarrow \gammap$ in the continuum limit.

Classically, the theory (\ref{eq:dirac_action}) has two global symmetries.  Under the vector rotation,
the spinors transform as:
\begin{equation}
\psi\rightarrow e^{i\alpha}\psi, \qquad 
\bar\psi\rightarrow \bar\psi e^{-i\alpha},
\end{equation}
and under the axial rotation, they transform as:
\begin{equation}
\label{eq:chiral_symmetry_global}
\psi\rightarrow e^{i\alpha\gammap}\psi, \qquad
\bar\psi\rightarrow \bar\psi e^{i\alpha\hat \gammap}\,,
\end{equation}
where $\alpha$ is some space-time constant.
It is well-known that the axial symmetry 
suffers from a quantum anomaly when the path integral is properly
regularized in the continuum.  On the lattice, this anomaly 
appears precisely due to the fact that the axial rotation performed 
above does not leave the fermion measure
invariant.  Suppose $(u_1, u_2, \dots, u_d)$ forms an orthonormal spinor basis 
of dimension $d$, and expand the fermion fields as follows:
\begin{equation}
\psi=\sum_{n=1}^{d}a_n u_n, \qquad \bar\psi=\sum_{n=1}^{d}\bar b_n u_n^\dag~.
\end{equation}
The fermion measure is then just given by:
\begin{equation}
\label{vectormeasure}
\mD\bar\psi\mD\psi=\prod_{n=1}^{d}\ud \bar b_n\prod_{n=1}^{d}\ud a_n.
\end{equation}
Under the axial rotation of eqn.~(\ref{eq:chiral_symmetry_global}):
\begin{equation}
\label{axialtransform}
\psi'=\sum_n a_n'u_n=e^{i\alpha\gammap} \sum_n a_n u_n,\qquad
\bar\psi'=\sum_n \bar b'_n u_n^\dag =\sum_n \bar b_n u_n^\dag e^{i\alpha\hat\gammap}~,
\end{equation}
we find that to leading order in $\alpha$:
\begin{gather}
a'_n=a_n+i\alpha\sum_m a_m (u_n^\dag, \gammap u_m),~~
\bar b'_n=\bar b_n+i\alpha\sum_m \bar b_m (u_m^\dag\hat\gammap, u_n).
\end{gather}
Therefore, the Jacobian of the infinitesimal field transformation (\ref{axialtransform}) is given by:
\begin{equation}
\label{abc}
\begin{split}
J^{-1}=&\left[1+i\alpha\sum_n (u_n^\dag\hat \gammap, u_n)\right]
\left[1+i\alpha\sum_n(u_n^\dag, \gammap u_n)\right] \\
=&(1+i\alpha\Tr\hat\gammap)\cdot(1+i\alpha\Tr\gammap)=1+i\alpha\Tr\hat\gammap.
\end{split}
\end{equation}
We have used the fact that $\Tr \gammap=0$.  The trace of $\hat\gammap$,
on the other hand, does not vanish in topologically
nontrivial gauge-field backgrounds. 
The properties  $\hat\gammap^\dag=\hat\gammap$ and $\hat\gammap^2=1$ alone
are sufficient to deduce that the eigenvalues of $\hat\gammap$ are $\pm 1$.
Therefore, the trace  appearing in (\ref{abc}), 
\begin{equation}
\label{index}
\Tr \hat\gammap=n_+-n_-,
\end{equation}
must be an integer\footnote{It must be an even integer since $n_++n_-=d$ is even.}.
Here $n_\pm$ are the dimensions of the invariant vector spaces
spanned by the eigenvectors of $\hat\gammap$ of $\pm 1$ eigenvalues
respectively.  Obviously $n_+ +n_-=d$. This is the lattice version of the index
theorem in the GW formalism.

There is a significant amount of literature\cite{Hasenfratz:1998ri,Fujikawa:1998if,F2,Adams:2000yi,Adams:2001jd} on many aspects of the expression (\ref{index}) that will not be discussed here. For now, we only mention that one finds, using the definition of $\hat{\gamma}_5$ and the explicit expression  for $\uD$, expanded in the continuum limit:\footnote{We use ``Tr" to denote trace over spacetime, gauge, and representation indices, while ``tr" only includes spinor and representation trace.}
\begin{equation}
\label{eq:tr_gamma_hat}
\tr (\hat\gammap_{xx})=-\frac{T(\cal{R})}{32 \pi^2}\, \epsilon_{\mu\nu\alpha\beta}
F^{\mu\nu}_aF_a^{\alpha\beta}(x)+\textrm{~higher dimensional terms}\dots\,,
\end{equation}
  for 4d Dirac fermions transforming in a representation $\cal{R}$ of $SU(N)$, with quadratic Casimir $T(\cal{R})$ ($1/2$ for the fundamental), or: 
  \begin{equation}
 \label{2dtracegammahat}
\tr (\hat\gammap_{xx})=-\frac{1}{2\pi} \epsilon_{\mu\nu}
F^{\mu\nu}(x)+\textrm{~higher dimensional terms}\dots\,,
\end{equation}
  for a 2d Dirac fermion of unit charge   under a $U(1)$ gauge group; in both expressions, 
 $x$ denotes an arbitrary lattice site. The  expressions (\ref{eq:tr_gamma_hat}), (\ref{2dtracegammahat}) show that 
 the usual index theorem of the continuum is recovered.

To summarize this Section, the advantage of the GW formalism is that it eliminates the fermion doubling problem and at the same time defines an exact chiral symmetry on the lattice, which reduces to the chiral symmetry (\ref{chi1}) in the continuum limit.  
Consequently, fermions with opposite ÒGW chiralitiesÓ can be disentangled in the action, allowing the definition of ``chiral partition functions," which play an important role in our further discussions. 

\subsection{Problems with  Ginsparg-Wilson fermions in chiral gauge theories   }
 \label{GWproblems}
 
Sadly, the beauty of the GW formalism is not the end of the story of defining chiral gauge theories and troubles arise the moment we consider dynamical gauge fields. The discussion of Section \ref{ambiguousphase} about the phase ambiguity of chiral partition functions finally becomes relevant. Again, consider the partition function:
 \begin{equation}
\label{GW5}
Z=\int D\bar\psi_L D\psi_L \exp\left( \bar\psi_L \uD\psi_L \right),
\end{equation}
but let now  $\bar\psi_L$ be the ÒGW chiral fermions,Ó as defined in (\ref{GW41}). To define $Z$, one chooses sets of orthonormal basis for both $\psi_L$ and $\bar\psi_L$, denoted as $\{u_i\}$ and $\{v_i\}$, respectively, and then evaluates $Z $ as $\det(v_i^\dag \uD u_i)$ (see also Section \ref{splitZ} for more details). As discussed in Section \ref{ambiguousphase}, the freedom of rotating either of the two basis by a unitary matrix renders the phase of $Z$ ambiguous.

What is changed in the GW formalism is that with dynamical gauge fields, one no longer has the option to just fix a set of ${v_i}$ and then stick to them. This occurs because the definition of $\bar\psi_L$ involves $\hat\gamma_5$, which depends on the Ginsparg-Wilson operator $\uD$, which in turn depends on the gauge field background. When the gauge field varies, the subspace in which $\bar\psi_L$ lives rotates  along. The same set of $v_i$'s inevitably fails to span it. It becomes mandatory to choose a different orthonormal basis $\{v_i\}$ for every gauge field configuration, and all of them are subject to arbitrary unitary rotations. One finds, in the Ginsparg-Wilson case, not a single ambiguous phase but a $U(1)$ bundle fibered over the entire gauge field configuration space, or more rigorously speaking, a $U(1)$ bundle fibered over the gauge field configuration space modulo all the gauge transformation---since the phase of $Z$ should be at least gauge invariant (for an anomaly-free theory) to be considered interesting. Finding a particular way of fixing the phase of $Z$ for all gauge field backgrounds is equivalent to finding a global section of this $U(1)$ bundle. It is not obvious whether such a global section exists and whether it is unique when it does. The problem of determining the phase of $Z$ is often referred to as the ``fermion measure problem" in lattice chiral gauge theories.

 Although there is no natural choice of a connection on this $U(1)$ bundle, there exists  a natural (i.e.,   independent of the basis vectors) definition of its curvature, solely determined in terms of the GW operator $\uD$. This  
 shows that the phase of the partition function can not be chosen at random---in particular, if  the phase is  taken to be gauge-field independent, the partition function (\ref{GW5}) is not even gauge invariant, even in the anomaly-free case.\footnote{Discussing  this topic in detail is beyond the scope of this review. We refer the reader to the original papers\cite{Neuberger:1998xn,Luscher:1998pq,Lu2,Luscher:1998du} or the reviews\cite{Golterman:2000hr,Luscher:2000hn,Neuberger:2001nb}.  For an illustration (within  the Wilson-line subspace of the gauge-field configuration space) of the topological obstruction to defining $Z$ in the anomalous case and an explicit  construction of the phase of $Z$ in the anomaly-free case,  see\cite{Poppitz:2007tu}.} 
 Integrating this curvature over any nontrivial cycles in the gauge field configuration space, 
Neuberger\cite{Neuberger:1998xn} and L\" uscher\cite{Luscher:1998pq,Lu2,Luscher:1998du}  found that it vanishes precisely when the gauge-anomaly cancellation condition is obeyed. This result indicates that in the anomaly-free case a global section can be found.  A proof that  a global section---one that ensures gauge invariance, locality, and smoothness of $Z$ as a function of the gauge background---can be found and is unique, up to usual irrelevant counterterm ambiguities, in the case of an anomaly-free abelian gauge group, was given in\cite{Luscher:1998du}. 
 Despite L\" uscher's existence proof, a practical way of implementing the theory on a computer is still missing; however, see the work of refs.\cite{Kadoh:2007xb,Kadoh:2007wz} on $U(1)$ and $SU(2) \times U(1)$ chiral gauge theories.
At the moment of writing, a generalization of the proof to general anomaly-free nonabelian chiral gauge theories is not known.
  
  Apart from the fermion measure problem, there are other issues with the formulation of lattice chiral gauge theories with Ginsparg-Wilson fermions  by the path integral  (\ref{GW5}):
\begin{enumerate}
\item The issue of summation over different topological sectors. This issue arises because in different topological backgrounds the dimensionality of the fermion path integral in (\ref{GW5}) changes. This is because 
the dimensionality of the $\bar\psi_L$ space is equal to tr$\hat{P}_+$, which has different values in different topological sectors (on the other hand, the dimensionality of $\psi_L$ space is gauge-background independent, equal to tr$P_+$). What is not known is how to determine the relative normalization of the contribution of different topological  sectors to expectation values of observables via  (\ref{GW5}).  This is not an issue in vectorlike theories, where the dimension of the $\bar\psi_{L+R}$ space is   tr$\hat{P}_+ +$tr$\hat{P}_- =$ tr${P}_+ +$tr${P}_-$ and is thus equal to the dimensionality of the $\psi_{L+R}$ space for any gauge background. In this context, we 
note the proposal of\cite{Suzuki:2000ku} to determine the absolute value of the normalization factors, but not their phase, starting from a real-representation vectorlike theory (however, it is not clear if this normalization is consistent with cluster decomposition). If the ``decoupling of mirror fermions" approach discussed in this review  is successful, this  normalization issue is resolved automatically,  since the measure is that of the vectorlike theory.

\item The unwanted-CP-violation problem. Under the condition 
of locality, the gauge-field dependent projectors $\hat{P}_\pm$  have an asymmetry between 
fermions and anti-fermions. As this asymmetry provides the origin of fermion number violation 
in the chiral theory, it is a welcome feature. However, at the same time it breaks CP\cite{Hasenfratz:2001bz,Fujikawa:2002is,Fujikawa:2002vj}. This CP violation afflicts high-momentum modes and might be (naively) expected to disappear in the continuum limit; furthermore, it   may   also show up in instanton-induced  fermion-number violating amplitudes\cite{Hasenfratz:2001bz}. We stress  that the effect of this CP violation in the quantum continuum limit is not completely understood and that the CP violation issue also afflicts the mirror-decoupling approach (if standard projectors and CP are used), where its implications await further study.
Note that  the above remarks refer to the continuum CP transformation: recently ref.\cite{Igarashi:2009kj} argued that in the trivial topological sector it is possible to define a lattice-modified version of CP, (which, like the modified chiral symmetry, reduces to the usual continuum CP-transform), under which the chiral action (\ref{GW5}) (and measure) is invariant; see also\cite{Hasenfratz:2007dp,Cundy:2010pu}. 
\end{enumerate}

\section{Chiral from vectorlike?}
\label{chiralfromvectorlike}

\subsection{The idea of decoupling the mirror fermions}
\label{ideaofdecoupling}
  
Our approach to tackle the long-standing problem of defining lattice chiral gauge theories is to try to work around the toughest part. Realizing that defining the measure of an arbitrary chiral gauge theory   explicitly is beyond our reach, we propose to start with a vectorlike theory instead.\footnote{We should note right away that this is not a completely new idea. However, when coupled with exact lattice chirality, it acquires  many new 
aspects, see Section \ref{preginspargwilsoncomparison} for discussion and references.}
The measure and the action of a vectorlike theory can be unambiguously defined. 

Thus, the vectorlike theory we consider should be described by an action $S$ which splits into two separate Òchiral partsÓ:
 \begin{equation}
 \label{EP1}
S=S_{\textrm{light}}+S_{\textrm{mirror}}~.
 \end{equation}
 The fields that participate in the two parts of the action will be called
 the ``light" and ``mirror" fields, respectively. The light  sector describes a chiral gauge theory that is the target theory we wish to obtain in the end, while the mirror sector contains all the fermions with the wrong chiralities. As explained  earlier, such a clear separation is only possible on the lattice using the GW formalism; hence, the so-called chiral fermions refer to the GW chiral fermions. 
 
 The hope  is to design the mirror sector in such a way that all the mirror fermions of wrong chiralities contained there become heavy and  disappear in the IR limit. If the gauge symmetry remains unbroken by the mirror dynamics, we would obtain a chiral gauge theory described by the action $S_{\textrm{light}}$.
 Since the measure is that of the vectorlike theory, we never  have to worry about choosing the ambiguous phase, thus circumventing the difficulty of defining the chiral fermion measure explicitly. We call this idea ``decoupling of the mirror fermions."
 
To better explain what we are proposing, let us first work in continuum terms and explain the desired features of the mirror dynamics. 
 We will use the chiral $SU(5)$ theory already discussed in the Introduction to illustrate the idea.
We use two-component Weyl-spinor notation to describe the desired ``light" fermions of the target $SU(5)$ theory:
\begin{equation}
\label{su5-1}
{\rm light:} ~~ \psi^i_\alpha \sim {\bf 5^*} ,~~ \chi_{ij \; \alpha}  \sim {\bf 10}~,~~\zeta_\alpha \sim {\bf 1}~,
\end{equation}
where  $i$ denotes an $SU(5)$ (anti-)fundamental index and $\alpha = 1,2$---an $SL(2,C)$ index. We also introduce the   ``mirror" partners of (\ref{su5-1}): 
\begin{equation}
\label{su5-2}
{\rm mirror:} ~ ~ \eta_{i \alpha} \sim {\bf 5} ,~~ \rho^{ij}_\alpha  \sim {\bf 10^*}~, ~~\xi_\alpha \sim {\bf 1^*}~.
\end{equation}
In this notation a Dirac mass term for the ${\bf 5+5^*}$ would be of the form $m \psi^{i \alpha} \eta_{i \alpha} + {\rm h.c.}$. We also introduced a gauge-singlet  Dirac fermion, with Weyl components  $\zeta, \xi$. This is a field whose $\xi$ component will play  an important role in the strong mirror dynamics (its $SU(5)$ representation was denoted by $\bf 1^*$ simply to distinguish from its partner).\footnote{An entire gauge-singlet Dirac multiplet was added to make sure the fermion representation (\ref{su5-1}, \ref{su5-2}) is vectorlike (even though this is not strictly  necessary in the singlet sector).} As already stated, we imagine that on the lattice the chiral components are defined as appropriate for   GW fermions (\ref{GW41}). 

Now, to decouple the mirrors (\ref{su5-2}), we add non-gauge interactions involving only the mirror fields. Since the mirror fermions are in a chiral representation, there are no mirror-fermion gauge-invariant bilinears that one can write down. However, three gauge-invariant four-fermion interactions can be written 
down. In group theory terms, these invariants are $\bf 5$-$\bf 10^*$-$\bf 10^*$-$\bf 10^*$, $\bf 1$-$\bf 5$-$\bf 5$-$\bf 10^*$, and  $\bf 1$-$\bf 1$-$\bf 1$-$\bf 1$. More explicitly, the mirror action is:
\begin{equation}
\label{su5-3}
S_{\textrm{mirror}} = \lambda_1 \eta^\alpha_i \rho_\alpha^{ij} \rho^{\beta \; kl} \rho_\beta^{mn} \epsilon_{jklmn} +
 \lambda_2 \xi^\alpha \eta_{\alpha\; i} \eta^\beta_j \rho^{ij}_\beta + \lambda_3 \xi^\alpha \xi_\alpha \xi^\beta \xi_\beta~ + {\rm h.c.},
 \end{equation}
 where we omitted the mirror kinetic terms, which have the standard form. 
The reason for introducing the  mirror interactions $\lambda_{1,2,3}$ is to decouple the mirror fermions while preserving the gauged
$SU(5)$ symmetry. This might be possible  if the   four-fermi interactions (\ref{su5-3}), when  
taken strong, lead  to the formation of $SU(5)$ invariant (or vectorlike) mirror composite states, which can  acquire mass without breaking $SU(5)$. 
For example, a possible composite of the mirror fermions is the $\eta\eta\rho$ (${\bf 5}$-${\bf 5}$-${\bf 10^*}$) invariant appearing in (\ref{su5-3}).  This composite state can   acquire a large Dirac mass by pairing with the singlet mirror field $\xi$ and can thus decouple  from the low-energy physics without breaking the $SU(5)$ symmetry. Similarly, at strong coupling, all mirror fermions are expected to be   bound in massive  singlet or vectorlike composites and decouple from the infrared dynamics. 

Thus, the desired spectrum of the theory  consists of heavy mirror states, with mass of order of the ultraviolet cutoff (see  Section \ref{toymodel}). The $SU(5)$ gauge interactions are only a spectator to the strong mirror dynamics. The gauge interactions couple to both the light and mirror states, but the gauge coupling at the cutoff scale is expected to be weak and is not expected to cause the heavy mirror states to become light or the massless light states to become heavy (the latter are protected by their unbroken exact chiral symmetries). 
Thus, at scales below the cutoff, the spectrum and the global chiral symmetries of the light fermions would be exactly  the ones of the desired chiral gauge theory.

There are several pieces of evidence that the previous two paragraphs are not  complete fantasy. 

The first---and most pertinent to our approach---is the existence of strong-coupling symmetric phases of lattice four-fermi or Yukawa models. In the following Section \ref{toymodel}, we consider a simple toy example of a strong-coupling symmetric phase in a theory with naive lattice fermions. In this phase, the spectrum has only massive states and an unbroken global symmetry. The nature of similar strong-coupling symmetric phases in theories  with GW fermions is not yet completely understood and is  the subject of later Sections.

The second---less directly relevant to our approach, but nonetheless tantalizing---motivation is provided by the  ${\cal{N}}=1$ supersymmetric examples where chiral and vectorlike gauge theories  are related by Seiberg dualities. There, a theory with charged matter content in vectorlike representations---but with some chiral couplings to singlets in the superpotential, not unlike (\ref{su5-3})---is argued to have an equivalent infrared description in terms  of a theory with  chiral matter content. This example shows that, at least in the supersymmetric ``theory space," gauge theories with chiral and vectorlike matter content may be  related. 
The first example of chiral-nonchiral Seiberg duality was found by Pouliot\cite{Pouliot:1995zc} and was later generalized by Pouliot and Strassler\cite{Pouliot:1995sk}. They considered 
 the  vectorlike ${\cal{N}}=1$ supersymmetric $SO(8)$ gauge theory with one chiral superfield in the spinor and $N$ chiral superfields in the vector representation (both are real, eight-dimensional representations),
 along with some gauge-singlet chiral superfields with chiral Yukawa couplings.
They argued that this vectorlike theory has a dual  infrared description in terms of a
  chiral  $SU(N)$ theory ($6\le N\le 16$) with a symmetric tensor and $N+4$ antifundamental chiral multiplets without any superpotential. The two dual theories---the chiral and the vectorlike one---flow to an interacting conformal field theory. The  chiral primary  operators  of the two theories and their scaling dimensions (which can be calculated in supersymmetry  even at strong coupling)  can be mapped to each other and  the duality map can be  shown to pass some rather intricate consistency checks. 
Furthermore, as in Seiberg duality between vectorlike theories, a free magnetic phase\footnote{See the already mentioned\cite{Strassler:2001px} for references.} occurs for $N\ge 17$---where the dual vectorlike $SO(8)$ theory loses asymptotic freedom. Thus, the asymptotically free chiral $SU(N-4)$ theory flows in the IR to a theory of free $SO(8)$ vectorlike quarks and gluons (and some gauge singlets), which can be viewed as the composite quasi-particles  of the  strongly-coupled chiral  theory.\footnote{
  It would be interesting to know if examples of chiral-nonchiral dual pairs with a free magnetic phase described by a chiral theory exist (this question  is not easy to answer, as there is no known  ``algorithm" for finding Seiberg duals). If so,   the   chiral theory would arise  as an effective weakly-coupled IR description of the vectorlike theory: a more ambitious goal than what we are trying to accomplish with the ``mirror decoupling" idea (as our aim is to only make the mirror fermions heavy without breaking the gauge symmetry, which is only a spectator to the strong mirror dynamics). In a chiral magnetic dual, the gauge fields and the fermions of the chiral theory would appear as composite objects of the vectorlike theory---and would thus offer an sintriguing possibility.}

 \subsection{A toy model of decoupling in strong-coupling symmetric phases.}
 \label{toymodel}
 
In this Section, we consider a simple toy model illustrating the idea of lattice strong-coupling phases in theories with multi-fermion interactions. The simple model below was invented solely for the purpose to give a short introduction to  strong-coupling symmetric phases in multi-fermion or Yukawa theories on the lattice. See\cite{Fradkin:1978dv,Forster:1980dg,Lang:1981qg,Lee:1987eg,Lee:1989mi,Hasenfratz:1988vc,Lee:1989xq,Aoki:1990at,Stephanov:1990pc} for  detailed studies of related models, where the entire phase diagram is also discussed; for brevity, we only focus on the strong-coupling symmetric phase as it is the one relevant for us.

We will consider our  lattice toy model in the Hamiltonian formulation (where space is a lattice, but time is continuous). The fields are 
 one-component fermion fields $\psi_a(x)$, $a=1, \ldots 4$, living on the sites $x$ of a lattice of any dimensionality. They obey the canonical anticommutation relations:
 \begin{equation}
 \label{commrel}
 \{ \psi_a(x), \psi^\dagger_b(x) \} = \delta_{ab} \delta_{xy}~, 
 \end{equation}
 and have a local 4-fermion interaction:
 \begin{equation}
 \label{4fermitoy}
 H_{int} = \sum\limits_{x, a} \lambda \left(\psi_a(x) \psi_b(x) \psi_c(x) \psi_d(x) \epsilon^{abcd} + {\rm h.c.}\right).
 \end{equation}
 There are also some hopping terms in the Hamiltonian, whose particular form is not important in what follows, say: 
 \begin{equation} \label{kinetictoy} H_{kin} = \sum\limits_{x,a, \hat{\mu}} \psi^\dagger_a(x+\hat{\mu}) \psi_a(x)\;+ \ldots ,\end{equation} 
 where $\hat{\mu}$ is a unit lattice vector in the $\mu$-th direction. The lattice spacing has been set to unity. The Hamiltonian preserves an $SU(4)$ global symmetry. The  analogy we want to keep in mind is that the 4-fermion Hamiltonian  (\ref{4fermitoy}) is similar to the mirror interactions in the $SU(5)$ theory (see 
 eqn.~(\ref{su5-3})) and that the global $SU(4)$ symmetry is the analog of the to-be-gauged $SU(5)$. 
 
We will   study  the limit $\lambda \gg 1$, when the four-fermi coupling is strong in lattice units. In this limit, the Hamiltonian of our toy model can be easily diagonalized, since to first approximation one can neglect the hopping term (\ref{kinetictoy}), decoupling all lattice sites  in (\ref{4fermitoy}). Thus, it is possible to find the eigenstates of the Hamiltonian (\ref{4fermitoy}) on every lattice cite and then build the ground state of the theory as a direct product of all single-site ground states. To diagonalize the single-site Hamiltonian, we denote
$\psi_b \rightarrow a_b$, $\psi^\dagger_b \rightarrow a^\dagger_b$, where $a_b$, $a^\dagger_b$ are four fermion creation and annihilation operators obeying the usual canonical anticommutation relations $ \left\{ a_b, a^\dagger_c \right\} =\delta_{bc}$. The single-site Hamiltonian is a simple quantum mechanics problem of four fermions with an interaction that preserves fermion number $F$ modulo 4 and an $SU(4)$ global symmetry:
\begin{equation}
\label{toysinglesite}
H_0 = \lambda \left( a_a a_b a_c a_d + a^\dagger_a a^\dagger_b a^\dagger_c a^\dagger_d \right)\epsilon^{abcd} ~.
\end{equation} 
\begin{table}[h]
\tbl{The $2^4$ states of the single-site four-fermi Hamiltonian (\ref{toysinglesite}).}
{\begin{tabular}{@{}ccc@{}} \toprule
$2^4$ states & $SU(4)$ rep. & F \\ \hline
$| 0 \rangle$ & $|\bf{1} \rangle$ & 0 \\
$a_b^\dagger | 0 \rangle$ & $|\bf{4} \rangle$ & 1 \\
$a_b^\dagger a_c^\dagger | 0 \rangle$ & $|\bf{6} \rangle$ & 2 \\
$a_b^\dagger a_c^\dagger a_d^\dagger| 0 \rangle$ & $|\bf{4^*} \rangle$ & 3 \\
$a_1^\dagger a_2^\dagger a_3^\dagger a_4^\dagger| 0 \rangle$ & $|\bf{1^\prime} \rangle$ & 4 \\
\end{tabular} \label{ta1}}
\end{table}
The Hilbert space of the single-site problem has $2^4$ states which can be decomposed as $SU(4)$ representations of given fermion number $F$, with $|0\rangle$ denoting the Fock vacuum, as shown in Table \ref{ta1}.  Conservation of $F$(mod)4  implies that only $H_0$ only has nonzero matrix elements between the states $| \bf{1} \rangle$ and $|\bf{1^\prime}\rangle$. Thus,  we find that the state $| \bf{1} \rangle - |\bf{1^\prime}\rangle$ has energy $-\lambda$, the state $| \bf{1} \rangle + |\bf{1^\prime}\rangle$ has energy $\lambda$, while the states $| \bf{4} \rangle$, $| \bf{6} \rangle$, and $| \bf{4^*} \rangle$ have zero energy. Hence, in the $\lambda \rightarrow \infty$ limit, the ground state of the single-site Hamiltonian is unique and is an $SU(4)$ singlet. The energy levels of $H_0$ are shown on Fig.~1. 
\begin{figure}[h]
\label{toyfigure}
\centerline{\includegraphics[width=7cm]{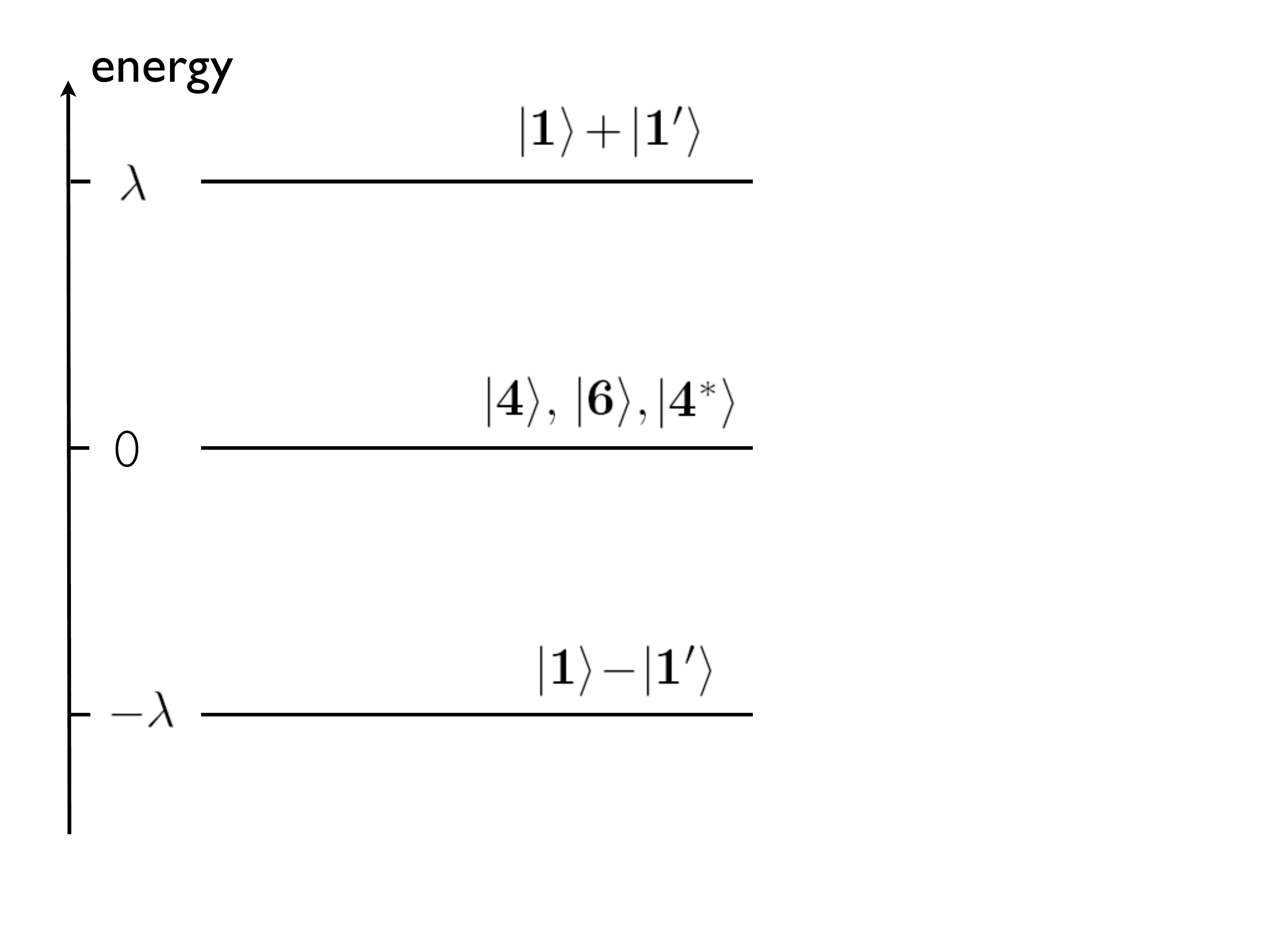}}
\vspace*{8pt}
\caption{Spectrum of the single-site hamiltonian (\ref{toysinglesite}) of the $\lambda \rightarrow \infty$ limit our toy model (\ref{4fermitoy}, \ref{kinetictoy}). The ground state is unique, has a gap $\lambda$ in lattice units, and preserves $SU(4)$. Taking hopping terms into account causes massive excitations to propagate between adjacent sites on the lattice. }
\end{figure}

We can now build the $\lambda \rightarrow \infty$ ground state of the entire lattice theory by simply taking the product of the $| \bf{1} \rangle - |\bf{1^\prime}\rangle$ ground states of every lattice site. This state is  unique and is an $SU(4)$-singlet.
 The name ``strong-coupling symmetric phase" should be clear now: the ground state of the four-fermi theory at $\lambda \rightarrow \infty$ is an $SU(4)$ singlet and all excitations are gapped, with mass gap proportional to $\lambda$ in lattice units. Taking hopping into account can be organized in a  $1/\lambda$ strong-coupling expansion; it causes the localized states to propagate from site to site, leading to propagating states of mass $\sim \lambda$. Since the strong-coupling expansion has a finite radius of convergence,  the picture of an $SU(4)$-singlet ground state and  only massive excitations is  a true representation  of the ground state of the finite-$\lambda$ theory, for sufficiently large $\lambda \gg 1$. A transition to a broken phase is expected to occur at some critical value $\lambda_c \sim 1$; this has been studied in the  literature quoted in the beginning of this Section.

The strong-coupling symmetric phase in this toy model has features similar to what we desire of the mirror dynamics   (\ref{su5-3}): for sufficiently strong mirror coupling it gives rise to only heavy mirror states and the to-be-gauged global $SU(4)$ symmetry is unbroken. The physics of the strong-coupling symmetric phase of this toy model is thus ``trivial," as there are no states lighter than the ultraviolet cutoff and thus no nontrivial infrared physics. Upon gauging the $SU(4)$ symmetry, however, there would be nontrivial infrared dynamics---that of the unbroken pure gauge theory\footnote{For a recent study of the decoupling of scalars in a strong-coupling symmetric phase in a simple toy model with gauge fields, see\cite{Poppitz:2008au}.}.

In his context, we note that the question of the ``triviality" of four-fermi (or Yukawa) interactions in 4d is often raised when their strong-coupling symmetric phases are mentioned. In our  context, triviality---interpreted as the absence of any long-distance physics in the strong-coupling symmetric phase---is  to be advantageously exploited, since the goal of the strong mirror interactions interactions is to keep all mirror fermions at the cutoff scale.

For the mirror-decoupling approach to chiral lattice gauge theories, the most important and still not completely answered question is whether models with lattice four-fermi mirror interactions formulated with Ginsparg-Wilson fermions have strong-coupling symmetric phases which share the same desired properties: an unbroken global symmetry {\it and} no massless mirror states. We already noted that using GW fermions allows us to separate the light and mirror modes in an unambiguous way while preserving all the chiral symmetries of the desired target theory. However, the dynamics of GW fermions is  more complicated than that of the simple 4-fermi model with local interactions (\ref{4fermitoy}). In particular, a strong-coupling expansion that preserves the chiral symmetries appears to be difficult to perform, because of the exponential-only\cite{Hernandez:1998et,Neuberger:1999pz} locality (which holds in the ``admissible" part of the gauge-field configuration space) of the exactly chiral lattice fermions. Thus, the numerical study of the strong-coupling phases appears the only way to proceed. We will describe the status of these studies in the following sections.

\subsection{Mirror global symmetries  and 't Hooft anomaly matching. }
\label{mirrorsymmetries}

In this Section, we explain why a necessary condition for the decoupling of the mirror fermions is that the mirror interactions explicitly break  all mirror global symmetries, including the ones that would have a gauge anomaly after gauging $SU(5)$.
To this end, note that in the limit when the $SU(5)$ gauge coupling and the interactions in (\ref{su5-3}) are turned off, the mirror theory has an $SU(5)\times U(1)^3$ global chiral symmetry. The $U(1)^3$ are phase transformations acting on each of the three mirror multiplets $\eta, \rho, \xi$. The  interaction terms in (\ref{su5-3}) explicitly break  $U(1)^3$, but preserve the to-be-gauged $SU(5)$.

The mirror theory is a theory with a chiral matter content and its unbroken chiral global symmetries have 't Hooft anomalies. 
The 't Hooft anomaly-matching argument is usually applied to asymptotically-free gauge interactions that become strong in the infrared.
However,  't Hooft anomaly matching also holds when strong non-gauge interactions on the lattice are considered (such as the ones in (\ref{su5-3})), when the theory is formulated with exactly chirally-symmetric GW fermions. As we  explain in   detail in  Sections \ref{transversality10}  and \ref{matching}, the reason is essentially that in  theories of chiral fermion formulated via GW fermions, the anomaly is independent of the coupling strengths and the details of the lagrangian (see  Section \ref{splitZ}).

For now, taking the anomaly-matching condition for granted,  note that the $SU(5)$ symmetry is respected by the strong mirror interactions in $S_{\textrm{mirror}}$, but since the mirror spectrum is anomaly-free, there is no $SU(5)^3$ 't Hooft anomaly to saturate. However, if one of the couplings $\lambda_{1,2,3}$ in (\ref{su5-3}) was set to zero, there would be an additional anomaly-free chiral $U(1)$
respected by the mirror interactions. This extra $U(1)$ would have some  't Hooft anomalies ($U(1)$, $U(1)^3$, or $U(1) SU(5)^2$, depending on which coupling was set to zero).  At zero mirror couplings, these anomalies are due to the elementary massless  mirror  fermions. The 't Hooft anomaly-matching conditions imply then that this anomaly has to be matched by the 
 spectrum of the strongly-coupled mirror theory---and thus light mirror states would be guaranteed to exist (whether they are bosons or fermions depends on the dynamics and can not be inferred from anomaly-matching arguments alone). Thus, we conclude that symmetry reasons alone require that all mirror global symmetries---except the one to be gauged---have to be explicitly broken, to prevent the appearance of massless mirror states required by 't Hooft anomaly matching. 
 
We stress that the absence of exact global chiral symmetries in the mirror theory is only a necessary condition---it does not  guarantee that the  mirror spectrum is necessarily free of massless fermions. The absence of anomalies to match simply means that any massless mirror states would have to come in anomaly-free representations under  the unbroken $SU(5)$---as would happen, for example, at weak mirror couplings $\lambda_{1,2,3}$. 
 However, our study  of   a two-dimensional model in Section \ref{symmetryvsexplicit} shows that,  at strong mirror coupling, the mirror spectrum saturates anomaly matching with the smallest required number of massless fermions and that massless fermions in anomaly-free representations do not occur when all mirror global symmetries are broken. Needless to say, we take this as an encouraging sign---but by no means a proof---of the viability of this approach.
  \begin{figure}[h]
\label{instantonfigure}
\centerline{\includegraphics[width=8cm]{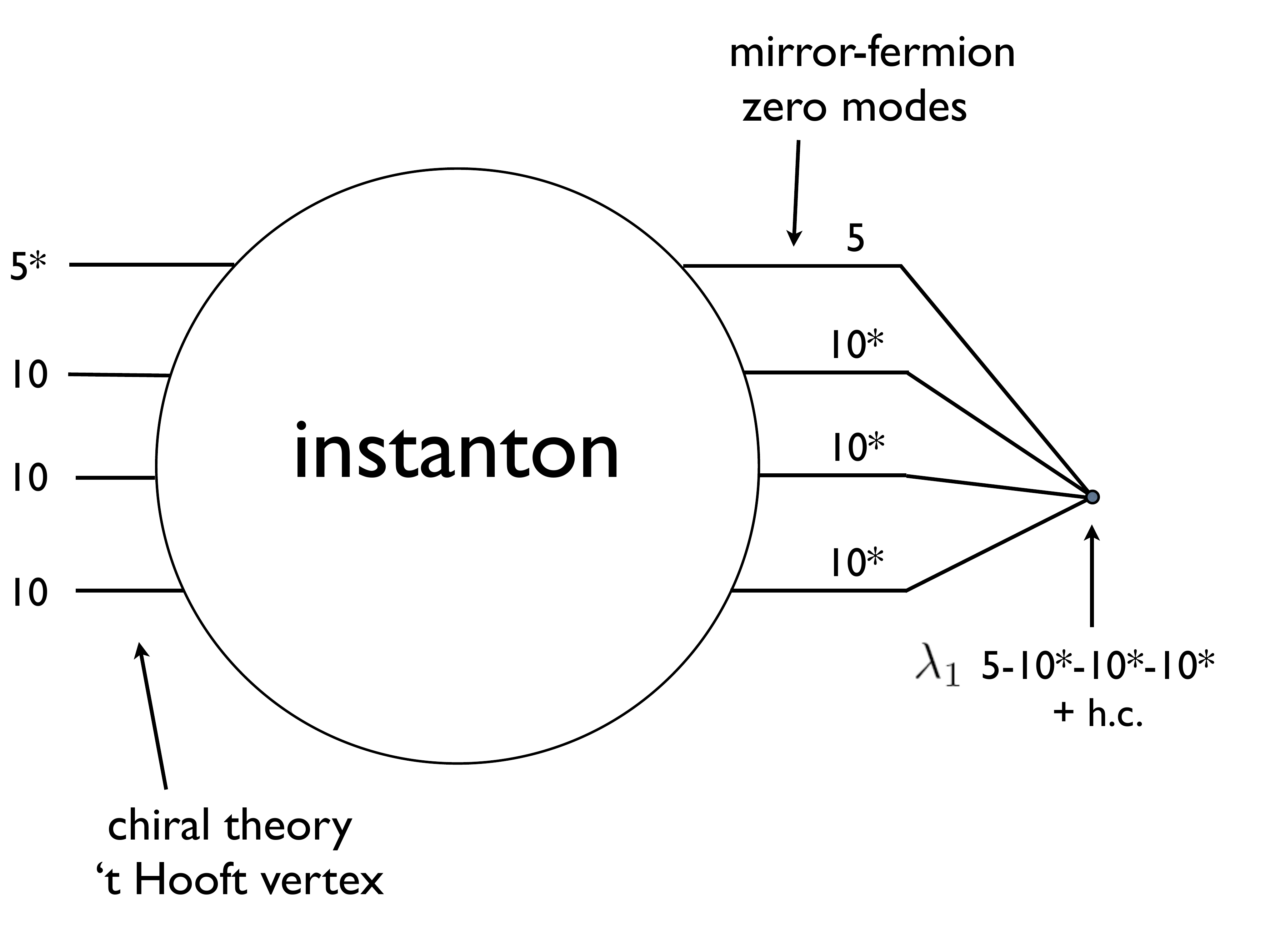}}
\vspace*{8pt}
\caption{In order to lift the mirror-fermion zero modes in an instanton background, the mirror interactions have to break the anomalous mirror global symmetries as well.}
\end{figure}

   To end this Section, we note that 
another way to see that mirror global symmetries that would be anomalous after gauging $SU(5)$ have to be explicitly broken, is to consider what would happen in a topologically nontrivial background, as discussed already in\cite{Eichten:1985ft}. In an instanton field, all chiral fermions, mirror and light alike, have zero modes as dictated by the index theorem. In the $SU(5)$ theory that we use as an illustration, this is shown on Fig.~2. In order to lift the mirror-fermion zero modes, one must include a mirror interaction that explicitly  breaks the anomalous mirror global chiral symmetry. In this case, the $\lambda_1$ term in (\ref{su5-3}) allows all mirror zero modes to be lifted and the target theory 't Hooft vertex to be reproduced.

Note the analogy with the Standard Model: in QCD, quark masses explicitly break the anomalous chiral symmetries violated in $SU(3)$-instanton backgrounds and allow heavy quarks to completely  decouple from the instanton vertex. In contrast, in the electroweak theory,  Yukawa couplings respect the anomalous $B+L$, violated in an $SU(2)_L$ instanton background, and thus heavy quarks do not decouple from the baryon-number violating 't Hooft vertex\cite{Banks:1992af}.
Thus, in order to be successful, our mirror-decoupling approach should resemble the case of QCD, where the anomalous chiral symmetry of the heavy quarks is explicitly broken.

   \subsection{Comparison to ``pre-Ginsparg-Wilson"  attempts of   mirror/doubler decoupling}
\label{preginspargwilsoncomparison}

This analysis of the toy model of Section \ref{toymodel} can be also performed for realistic models in  four spacetime dimensions (and can also be done in an Euclidean lattice formulation). The idea of the analysis is the same, but the calculations become more cumbersome in view of the proliferation of indices to keep track of.  

In fact, such a study was done by Eichten and Preskill\cite{Eichten:1985ft}
 for the $SU(5)$ model of Section \ref{ideaofdecoupling}. They studied the $SU(5)$ model with local four-fermion interaction  terms of the form  $ \lambda_1$$\bf 5$-$\bf 10^*$-$\bf 10^*$-$\bf 10^*$ 
 +$ \lambda_2$$\bf 1$-$\bf 5$-$\bf 5$-$\bf 10^*$  + h.c. (essentially the ones of (\ref{su5-3})  with $\lambda_3=0$). All fields were taken to be naive lattice Weyl fermions, in order to preserve the chiral symmetry.  
In this 1986 formulation it is the 15 doublers (on an Euclidean lattice) of the naive Weyl fermions that play the role of our ``mirrors."
Eichten and Preskill's ultimate goal---similar to ours---was to find a phase where the doublers decoupled and the light components remained exactly massless.

 The two couplings, $\lambda_1$ and $\lambda_2$, were motivated by: {\it a.)} having to lift the mirror-fermion instanton zero modes and {\it b.)} ensuring that the Euclidean path integral has a ``static limit." The first condition was already explained in the previous Section \ref{mirrorsymmetries}.
The second condition was the reason to add  the singlet fermion with coupling $\lambda_2$---if it is absent, the Euclidean path integral over fermions with kinetic terms put to zero vanishes, indicating that no static limit and no sensible strong-coupling expansion exist (in a Hamiltonian formulation this would mean that the ground state at each site is not unique---indicating that there would  be no stability with respect to hopping perturbations).

By an analysis which is ideologically  identical to the one of Section \ref{toymodel},
Eichten and Preskill showed that when $\lambda_1$ and $\lambda_2$ were taken strong in lattice units, the $SU(5)$ global symmetry remained unbroken, but all components of the Weyl fermions became heavy.
 Thus, while a strong-coupling symmetric phase exists, both the light fermions and the doublers are heavy, absent any additional couplings. They proceeded by adding two more four-Fermi terms with couplings $r_1, r_2$. These terms have the same form as in (\ref{su5-3}) but now include lattice derivatives 
designed to lift the light-doubler degeneracy. The hope was to  find a region in the enlarged $\lambda_{1,2}/r_{1,2}$ parameter space of the model, where there would be a strong-coupling symmetric phase---but now with doublers having a cutoff-scale mass and exactly massless light fields in a ${\bf 5 + 10^*}$ representation of the unbroken $SU(5)$. The model with these extra couplings was   analyzed in\cite{Golterman:1992yha}, using all available analytic techniques. The conclusion was that in all accessible corners of the phase diagram with unbroken $SU(5)$, either all fermions (light and doublers) were massless or all fermions obtained mass. 

While the study of ref.\cite{Golterman:1992yha} can not be considered as having given a proof,  we believe that the apparent failure of this approach can be traced back to the absence of lattice chiral symmetries. To this end, note that  in the Eichten-Preskill model, both the ``light" and ``mirror" (there: doubler) fermion components of the lattice Weyl fermions participate in the strong mirror interactions (recall that, in 1986, there was no known way to separate the chiral components from a lattice fermion field). While a mass splitting of the ``light" and ``mirror" (doubler)   fermions occurs through the $r_{1,2}$ derivative four-Fermi interactions mentioned above,  exactly massless light fermions and heavy mirrors could at best be expected to occur for special, finely tuned, values of the ``four-fermi Wilson" couplings $r_i$, as there was no lattice chiral symmetry protecting the light fields. A chiral symmetry  could only be expected to emerge at certain values of $r_i$.

We stress that, in addition---as well as prior, see\cite{Smit1} for references---to the work\cite{Eichten:1985ft}, there were many studies of mirror/doubler  fermion decoupling via strong Yukawa or four-Fermi interactions\cite{Golterman:1990zu,Golterman:1991re,Smit2,Smit1993,Golterman:1993th,Golterman:1994at} (strong Yukawa and strong four-Fermi interactions can be mapped to each other; this was part of the analysis of\cite{Golterman:1992yha}), reaching similar conclusions. The absence of lattice chiral symmetries was a feature of all these attempts. The lack of an exact chiral symmetry without doublers  makes it impossible to  split  a lattice Dirac fermion field into chiral components (or separate a naive Weyl fermion field into light and doubler parts).  As a result, any strong non-gauge interaction   inevitably  couples to both the ``mirror/doubler"  and   ``light" components of the  fermion. The analysis of the strong dynamics---which is clearly not purely ``mirror" anymore---of the chiral symmetries and their realization then becomes ambiguous. Most importantly, these formulations lacked a manifest symmetry explaining why some chiral fermions should stay exactly massless while others could obtain mass; instead, chiral symmetries distinguishing ``light" from ``mirror" modes and protecting the ``light" modes were expected to somehow emerge  at special values of the couplings ($\lambda_{1,2}$/$r_{1,2}$ in the above example); see\cite{Aoki:1991es} and references therein for related studies in Yukawa models. 
  
In contrast to the earlier approaches, the advent of  exact lattice chiral symmetry in 1997 allows the construction of lattice actions where only the mirror fields participate in the strong interaction (a  proposal along these lines was made\cite{Creutz:1996xc} in the finite domain-wall context, shortly before the advent of exact chiral lattice symmetry). The chiral symmetries of the mirror theory, their 't Hooft anomalies, and their realization can now be unambiguously defined and studied.  The unbroken global  chiral symmetries  are exactly those of the target continuum chiral gauge theory and 
are manifest at finite lattice spacing and volume (as argued in\cite{Bhattacharya:2006dc}), leading to an elegant formulation of the mirror-decoupling idea.

\subsection{The big picture}
\label{generaldiscussion}

It is now straightforward to formulate  mirror decoupling via Ginsparg-Wilson fermions, for example for the $SU(5)$ model considered in Section \ref{ideaofdecoupling}. The action consists of kinetic terms  for the vectorlike
fermions transforming in the $\bf{5^*}$ and $\bf{10}$ representation, the singlet fermion, plus the mirror action (\ref{su5-3}). A light-mirror split of the lattice fermion action:
\begin{equation}
S_{fermion} = S_{light} + S_{mirror}~,
\end{equation}
is possible because of the existence of lattice chirality, which allows  to split a vectorlike fermion in chiral components at finite lattice spacing.   The aim of the  discussion here is to pose the various questions---answered or not---that arise in this approach for general models.
 
 The fermion partition function in a given gauge background   can be written as follows: 
\begin{equation}
\label{vectorz}
Z_{vector}[A] = \int d \psi d \bar\psi e^{S_{light}[A] + S_{mirror}[A]}~,
\end{equation}
where $d \psi d \bar\psi$ is  the well-defined fermion measure of the vectorlike theory, see eqn.~(\ref{vectormeasure}). Exact lattice chirality allows (see Section \ref{splitZ}) the partition function $Z_{vector}$ to be split, schematically:
\begin{equation}
\label{vectorzsplit}
Z_{vector}[A] = Z_{light}[A] \times Z_{mirror}[A]~
\end{equation}
into light and mirror partition functions.
The light partition function is that of the chiral $\bf{5^*}$ and $\bf{10}$ gauge theory, while the mirror partition function includes the mirror fermions plus their interactions.  
An explicit form of  (\ref{vectorzsplit})  in the 2d models whose dynamics we have investigated will be given Section \ref{splitZ}. 

The split form of the partition function (\ref{vectorzsplit})   is useful, because it allows us  to focus attention on  the non-gauge mirror dynamics described by $Z_{mirror}[A]$, first without gauge fields (or  with only perturbative gauge fields turned on). An investigation of the mirror dynamics without dynamical gauge fields is a sensible first step in understanding the possibility of mirror-fermion decoupling.  

\subsubsection{Questions about the mirror dynamics}
\label{questions}

 As already stressed, the question of defining the fermion measure, and thus of the chiral theory partition function, does not appear here. However, the price to pay is the introduction of strong mirror dynamics, about which many  questions arise:
\begin{enumerate}
\item The multi-fermi (or Yukawa) interactions in the mirror theory formulated with GW fermions are only exponentially local, subject to the admissibility condition on the gauge field background\cite{Hernandez:1998et,Neuberger:1999pz}. Do strong-coupling symmetric phases exist?  
\item In typical models, there are many multi-fermi/Yukawa couplings in the mirror theory, e.g. (\ref{su5-3}), needed to break all necessary mirror global symmetries. Is there  a nontrivial phase structure as the ratios of these strong mirror couplings are varied? Is fine-tuning necessary to reach the strong-coupling symmetric phase?
\item Suppose one attempts to decouple an anomalous representation. Does the strong mirror dynamics avoid decoupling an anomalous representation  in a  symmetric phase (and thus creating  an inconsistent anomalous long-distance theory)? How?
\item The Yukawa/multi-fermi couplings in $S_{mirror}$ formulated with GW fermions  define a complex partition function. Is the long-distance theory unitary? 
\item Do  mirror fermions in anomaly-free representations decouple in the symmetric phase at strong coupling, as suggested by symmetry arguments? 
\item If the mirror dynamics works as hoped for, will this construction be useful for simulating any anomaly-free chiral gauge theory, in either 2d or 4 d?
\end{enumerate}

\subsubsection{Answers---known and unknown}
\label{answers}

We do not yet know the answers to all these questions. The answers below are based on symmetry arguments and evidence from studies of specific models:
 \begin{enumerate}
\item Strong-coupling symmetric phases with GW fermions have been shown to exist in both\cite{Giedt:2007qg}  2d  and\cite{Gerhold:2007yb,Gerhold:2007gx} 4d.
\item There is a nontrivial phase structure in the strong-coupling regime\cite{Giedt:2007qg} in the 2d model we have studied. No fine-tuning is needed to  reach the strong-coupling symmetric phase, as it exists for a wide range of couplings.
\item An anomalous representation does not decouple in the strong-coupling symmetric phase. A charged mirror fermion remains massless, realizing the minimal solution of the relevant 't Hooft anomaly matching condition\cite{Poppitz:2009gt}.
\item The result from (3) gives evidence for long-distance unitarity of the strong mirror interactions (as unitarity is crucial for 't Hooft anomaly matching).
\item We do not know the answer yet. However, numerical studies of mirror-decoupling of   anomaly-free 2d models are feasible and should provide evidence for (or against) decoupling in the near future.
\item This will depend on the severity of the sign problem. Based on continuum intuition, a first test of the method with dynamical gauge fields is likely to be possible in 2d chiral $U(1)$  or in the 4d  $j$$=$$3/2$ Weyl fermion  $SU(2)$ theory. 
\end{enumerate}

In the rest of this article, we will explain in more detail our answers above. Most of them were obtained from numerical simulations. This is because of 
the lack of an obvious controlled expansion to study the GW-fermion strong mirror dynamics.\footnote{Since, as already mentioned in Section \ref{toymodel}, the GW-fermion multi-fermion/Yukawa interactions are only exponentially local (subject to the admissibility condition\cite{Hernandez:1998et,Neuberger:1999pz}).}
 Due to the high cost of simulations with exactly chiral fermions, such studies are in their infancy. On the other hand, since the issues considered are ones of principle, analyzing two-dimensional models is a sensible first step. Showing that the ideas work in 2d will not prove that 4d chiral gauge theories can be similarly formulated (this will require 4d mirror studies), but the results are likely to provide insight into the relevant mirror dynamics. 
 
  Another appropriate simplification is to neglect the gauge field fluctuations, since gauge fields play  a spectator role to the strong mirror dynamics. If the strong non-gauge mirror dynamics  gives the mirror fermions mass of order the lattice cutoff, in a manner similar to the one   in the toy model,  asymptotic freedom of the gauge interactions and the exact chiral symmetry of the light  
fermions lead  us to expect that turning on dynamical gauge fields will not significantly affect the mirror dynamics or lift the massless modes. Questions about CP violation in topologically nontrivial backgrounds will have to be investigated, see Section \ref{GWproblems}.  
Including dynamical gauge fields would only be warranted after  the mirror decoupling  in zero gauge background is demonstrated.

 \section{Theoretical and Monte-Carlo studies of mirror decoupling via Ginsparg-Wilson fermions}
 \label{implementation}
\subsection{Strong-coupling phases and properties of chiral partition functions}
\label{chiralZ}
   
  Most of the  answers listed in Section \ref{answers} above were obtained in the framework of a toy 2d model, which we call    the ``1-0" model. This model is defined in Section \ref{10model}\footnote{The name ``1-0" model is reminiscent of a general class of anomaly-free $U(1)$ chiral gauge theories in 2d--the ``3-4-5", ``1-1-1-2", etc., models. Recall that the anomaly cancellation condition for a $U(1)$ chiral gauge theory in 2d states that the models should be ``Pythagorean", i.e. the charges $q_-$ of the left-handed and $q_+$ of the right-handed Weyl fermions should obey $\sum q_-^2 = \sum q_+^2$. Clearly a chiral gauge theory with ``1-0" charges of the left and right moving fermions (the ``light" or ``mirror" sectors of the 1-0 model) would be anomalous.} below.
  
We will also use the ``1-0" model to illustrate various theoretical results, which are more generally valid:  the splitting of a vectorlike partition function into chiral partition functions and an important result on the variation of chiral partition functions  with the gauge background---the ``splitting theorem"---a result without which many of the analytic and numerical studies would not be possible (see Section \ref{splitZ}).
   
\subsubsection{The ``1-0" model}
\label{10model}

The ``1-0" model is a 
  Yukawa-Higgs-GW-fermion model: a $U(1)$ two-dimensional lattice gauge theory with one Dirac fermion $\psi$ of charge 1 and a neutral spectator  Dirac fermion $\chi$. 
Considering this theory was motivated by its simplicity: it is the minimal Higgs-Yukawa-GW-fermion model in two dimensions with chiral Yukawa couplings. 
It holds the promise to yield, at strong Yukawa coupling, a strong-coupling symmetric phase and is, at the same time, amenable to numerical simulations not requiring the use of extensive computing resources. 

The   action of the ``1-0" model, with the gauge kinetic term omitted, is:
\begin{eqnarray}
S =  S_{light} + S_{mirror} \nonumber  
\end{eqnarray}
\begin{eqnarray}
S_{light}  =  - \left( \bar\psi_+ \cdot D_1  \cdot  \psi_+\right) - \left( \bar\chi_- \cdot  D_0  \cdot \chi_-\right) \nonumber   \end{eqnarray}
\begin{eqnarray}\label{10modelaction}
 S_{mirror} &=& S_\kappa - \left( \bar\psi_- \cdot D_1 \cdot  \psi_-\right) - \left( \bar\chi_+ \cdot D_0 \cdot  \chi_+\right) \nonumber  \\
&+& y \left\{ \left( \bar\psi_-  \cdot \phi^*  \cdot \chi_+ \right) + \left( \bar\chi_+ \cdot  \phi \cdot  \psi_- \right) \right\} \\   &+&  y h \left\{ \left( \psi_-^T \cdot \phi \gamma_2 \cdot  \chi_+ \right) - \left( \bar\chi_+ \cdot  \gamma_2 \cdot  \phi^* \cdot  \bar\psi_-^T \right)  \right\} \nonumber~.
\end{eqnarray}
 The chirality components\footnote{We use $\pm$, which is most commonly used in 2d (instead of $L/R$) to denote chiral fermion components.} for the charged and neutral barred fermions are defined by projectors including  the appropriate Neuberger-Dirac operators (charged $D_1$ and neutral $D_0$), for example $ \bar\psi_\pm = \bar\psi (1 \mp \hat{\gamma}_5)/2$. The brackets denote summation over the lattice sites as well as a spinor inner product.
The field $\phi_x = e^{i \eta_x}$, $|\eta_x| \le \pi$, is a unitary Higgs field of unit charge with the usual kinetic term:
\begin{eqnarray}
\label{Skappa}
S_\kappa =  \frac{\kappa}{2}\; \sum_{x} \sum\limits_{\hat{\mu}} \left[ 2 - \left(\; \phi_x^* \; U_{x, x+ \hat\mu} \; \phi_{x+\hat\mu} + {\rm h.c.}\; \right) \right]~,
\end{eqnarray}
where $U_{x, x+ \hat{\mu}} = e^{i A_\mu(x)}$. We will call the fermion fields that participate in the Yukawa interactions the ``mirror" fields---these are the negative chirality component, $\psi_-$, of the charged $\psi$, and the positive chirality component, $\chi_+$, of the neutral $\chi$---while the fields $\psi_+$ and $\chi_-$ will be termed ``light."

When the Yukawa couplings are absent, at $y=0$, (\ref{10modelaction}) is just the Schwinger model with a neutral Dirac spectator fermion and a unitary Higgs field of charge-1 and action $S_\kappa$. We will be interested in the small-$\kappa$ phase of the theory, when the unitary Higgs field fluctuates strongly and decouples at long distances (see\cite{Poppitz:2008au} for a recent study). 
We have  included  both Majorana ($yh$) and Dirac ($y$) gauge invariant Yukawa terms in the mirror action, in order to explicitly break all global symmetries of the mirror fermions (except of the $U(1)$ to be gauged, as explained in Section \ref{mirrorsymmetries}). 

When the unitary Higgs field acts essentially as a random variable (modulo correlations induced by small nonzero $\kappa$ and by  fermion backreaction),  the  similarity of the Yukawa interactions in (\ref{10modelaction})  with the multi fermi interactions of (\ref{su5-3}) is in that integrating out the essentially random Higgs field generates all multi-fermion interactions consistent with  the symmetries. 
Based on experience with strong-Yukawa expansions in theories with naive or Wilson fermions, it is expected  that in the large-$y$, fixed-$h$ limit, there is a symmetric phase. 

 The lattice action (\ref{toymodel}) completely defines the theory via a path integral over the charged and neutral fermion fields, the unitary higgs field, as well as the gauge fields. We will not perform the integral over the lattice gauge fields (but  will study  the variation of the partition function with respect to the small changes of the gauge background). Using this model, we hope to address some of the questions listed in Section \ref{questions}: for example, while (1), (2), (3), and (4) will be addressed, it is clear that we can not answer  question (5)  about decoupling an anomaly-free representation, since the mirror fermions in the ``1-0" model  are  in an anomalous representation. Thus, (5)---arguably one of the most important questions---is left for (near-) future study.

\subsubsection{Splitting the vectorlike partition function. The ``splitting theorem."
}\label{splitZ}

The partition function of the ``1-0" model---as well as that of any vectorlike theory---can be split into ``light" and ``mirror" parts in any  gauge background. 
To split the partition function, one uses the definite-chirality eigenvectors of $\hat\gamma_5$ and the  projectors $\hat{P}_\pm$ on the corresponding spaces:
\beqa
\label{uwbasis}
\hat\gamma_5 u_i &=& - u_i ~~,~~ ~~~~~\hat\gamma_5 w_i  = w_i~, \\
\hat{P}_- &=& \sum_i u_i u_i^\dagger ~~,~~ \hat{P}_+  =  \sum_i w_i w_i^\dagger = 1 - \hat{P}_- ~,
\eeqa
where we treat $u,w$ as columns and $u^\dagger, w^\dagger$ as rows. We also use  
the eigenvectors of $\gamma_5$ (the latter are independent of the gauge background) and the  projectors $P_\pm$:
\beqa
\label{vtbasis}
\gamma_5 v_i &=& v_i ~~,~~~~~~~ \gamma_5 t_i = - t_i~, \\
 {P}_+ &=& \sum\limits_i v_i v_i^\dagger ~~,~~ {P}_-  = \sum\limits_i t_i t_i^\dagger = 1 -  P_+ ~.
\eeqa
Using (\ref{uwbasis}), (\ref{vtbasis}), a general Dirac field $\Psi_x$,  can  be decomposed into chiral components $\alpha^+_i, \alpha^-_i$ via the $\gamma_5$ eigenvectors, while the conjugate spinor field $\bar\Psi_x$ is decomposed into chiral components $\bar{\alpha}^+_i, \bar{\alpha}^-_i$ using the $\hat\gamma_5$ eigenvectors, as follows:
\beqa
\label{psitoc}
\Psi_x  =  \sum_i \alpha^i_+ v_i (x) +  \alpha^i_- t_i(x)~,~~
\bar\Psi_x  =  \sum_i \bar{\alpha}^i_+ u_i^\dagger (x) +  \bar{\alpha}^i_- w_i^\dagger(x)~.
\eeqa

We now apply the split (\ref{psitoc}) to the fields $\psi$ and $\chi$ of the ``1-0" model, and note that
only the charged eigenvectors (of both light and mirror fields) depend on the  gauge background. The expansions (\ref{psitoc}) of the ``mirror" fields are explicitly given below:
\beqa
\label{evs1}
\chi_+  =  \sum_i \beta_+^i v_i  ~, ~~ \bar\chi_+ =  \sum_i \bar\beta_+^i u_i^\dagger[0] ~, ~~
\psi_-  = \sum_i \alpha_-^i t_i~,~~ \bar\psi_-   =  \sum_i \bar\alpha_-^i w_i^\dagger[A] . 
\eeqa
Clearly, expansions similar to (\ref{evs1}) hold for the ``light" fields as well:
\beqa
\label{evs2}
\chi_-  =  \sum_i \beta_-^i t_i  ~, ~~ \bar\chi_- =  \sum_i \bar\beta_-^i w_i^\dagger[A] ~,~~\psi_+  =  \sum_i \alpha_+^i v_i~,~~ \bar\psi_+ =  \sum_i \bar\alpha_+^i u_i^\dagger[A].
\eeqa
After substitution of (\ref{evs1}, \ref{evs2}), the partition function of the model (\ref{10modelaction}) splits  as follows:
  \beq
  \label{z01}
  Z[A; y, h] = Z_{light}[A] \times {1 \over J[A]} \times Z_{mirror}[A; y, h]~.
  \eeq
Here $Z_{light}[A] = \det || (u_i^\dagger[A] \cdot D[A] \cdot v_j) || \times$(similar determinant  for the neutral light spectator $\chi_-$) is the light sector partition function. The jacobian $J$ is a product of  jacobians  for the charged and neutral sectors; see Section 2.2. of\cite{Poppitz:2007tu} and Section \ref{mirrorpimunu} of  this paper for details. The mirror partition function is  given, more explicitly, by an integral over the charged mirrors ($\alpha_-, \bar\alpha_-$), neutral mirrors ($\beta_+, \bar\beta_+$), and unitary scalar field:
\beq
\label{mirrorZ1}
Z_{mirror}[A;y,h] = \int d^2 \alpha_- \; d^2 \beta_+ \; d \phi \; e^{- S_{mirror}}~,
\eeq
where the mirror action from (\ref{10modelaction}) is expressed in terms of the integration variables $\alpha_-, \beta_+$ and the eigenvectors via (\ref{evs1})  and $d \phi$ denotes a path integral over the phases of $\phi$.
The mirror-fermion integral  is thus a determinant which includes the kinetic term and  Yukawa terms from (\ref{10modelaction}) and the mirror partition function is the average of the determinant  over the random (in the disordered $\kappa \rightarrow 0$ phase) unitary field $\phi_x$.

When $A\ne 0$, the mirror partition function $Z_{mirror}$ (\ref{mirrorZ1}) depends on the gauge background through the operators entering $S_{mirror}$ (the Neuberger-Dirac operator and the associated projectors that appear in (\ref{10modelaction})) as well as through the gauge background dependence of the eigenvectors of $\hat\gamma_5$ used to split the partition function ($w_i[A]$, see (\ref{evs1})).
 
 The dependence of general chiral partition functions, with an arbitrary chiral action, (e.g., our $Z_{mirror}$) on the gauge background was studied in\cite{Poppitz:2007tu}. There,  an important technical result was derived: that under an arbitrary variation of the gauge background, the variations of the mirror partition function due to the variations of the eigenvectors (here: $w_i[A]$) and the operators entering the action (the Neuberger-Dirac operator $D[A]$ and the  projectors  $\hat{P}_\pm$ depending on it, collectively denoted by $O$ below) factorize. Explicitly, the ``splitting theorem" states that for an arbitrary variation $\delta$  of the gauge background,  defined by the first equality below,\footnote{A more general  chiral partition function depends on more than one set of $A$-dependent chiral eigenvectors, and a sum over their variations will appear on the r.h.s. of (\ref{deltaZchiral}). We stress again that (\ref{deltaZchiral}) is valid for arbitrary chiral mirror actions (naturally, (\ref{deltaZchiral}) gives the known result for the bilinear chiral actions appearing in (\ref{GW5})).} the chiral partition function changes as follows:
\beq
\label{deltaZchiral}
\delta \log Z_{mirror}[A] = \log {Z_{mirror}[A + \delta A]\over Z_{mirror}[A]} =  \sum_i (\delta w^\dag_i \cdot  w_i) +  \left< {\delta  S\over \delta O } \; \delta O\right> ~,
\eeq
where ``$\vev{\cdot}$'' denotes an expectation value calculated with the partition function $Z_{mirror}$.
 The splitting theorem\cite{Poppitz:2007tu} is valid for general chiral partition functions (e.g., the action does not have to be bilinear in the fermion fields) and is an important result, because:
 \begin{enumerate}
 \item It isolates the anomalies from the details of the interactions of the chiral theory and manifestly realizes on the lattice the idea that anomalies  are determined only by the representation  of the fields and not by the details of the Lagrangian.\cite{Poppitz:2007tu}
\item The splitting theorem  is indispensible\cite{Poppitz:2009gt} in the calculation of mirror chiral current-current correlators in a perturbative expansion in the gauge field.
 \end{enumerate}
In regard to (1) above, as already noted in Section \ref{GWproblems}, anomalies in chiral lattice gauge theories manifest themselves as  topological obstructions to smoothly defining the basis vectors (or more precisely, the ``measure current," $\sum_i (\delta w^\dag_i \cdot  w_i)$ in (\ref{deltaZchiral})) globally over the gauge field configuration space.\cite{Neuberger:1998xn}  This implies that 
  the  splitting of the partition function in the ``1-0" model into ``light" and ``mirror" sectors is singular,\cite{Poppitz:2007tu} since every sector is anomalous on its own (but the full vectorlike theory partition function is not singular as a function of the gauge background $A$). However, because the singularity is topological, much like the location of the Dirac string of a monopole, it can be moved around the gauge field configuration space. In particular, a perturbative expansion  around $A=0$ is insensitive to the singularity. Thus, the topological obstruction to making the splitting of the vectorlike partition function into ``light" and ``mirror" smooth over the entire gauge configuration space has no bearing on studies of the mirror spectrum at vanishing gauge background or infinitesimally close to it.

\subsubsection{The phase structure of the ``1-0" model at strong mirror Yukawa coupling}
\label{phasestructure}

The split  partition function (\ref{mirrorZ1}) allows the phase structure of the mirror sector of the ``1-0" model to be studied\cite{Giedt:2007qg}. The theory was considered in zero gauge background and  in the limit of strong Yukawa coupling ($y \rightarrow \infty$), where the mirror-fermion kinetic terms were dropped. As a function of $h$ (the ratio of Majorana to Dirac Yukawa couplings), a nontrivial phase structure was found. 
For $h \rightarrow 0$ and $h>1$ the theory is in a strong-coupling symmetric phase with only short-range correlations of the $U(1)$-charged order parameters. On the other hand, for $h \sim 0.7$, evidence for a transition to a ``broken" (more precisely, algebraically-ordered phase) was seen. The transition appears to be in the Berezinskii-Kosterlitz-Thouless universality class, as evidenced by  the change in the density of vortices\cite{Giedt:2007qg}. 
\begin{figure}
\begin{center}
\includegraphics[width=4in]{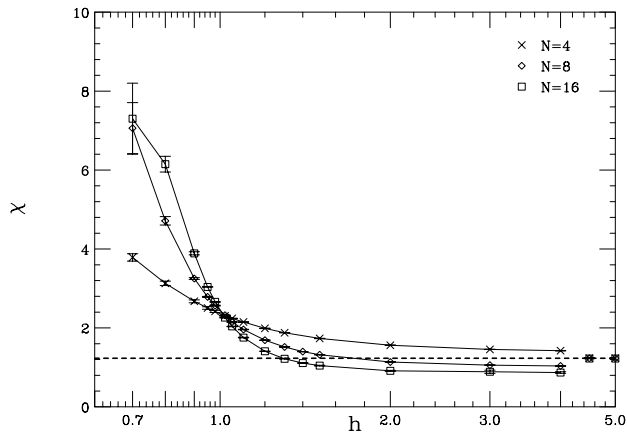}
\caption{Scalar-field susceptibility (\ref{chiscalar}) of the mirror sector of the ``1-0" model for $\kappa=0.1$, at $y\rightarrow \infty$. The dashed line indicates the susceptibilities for the pure $XY$ model with the same $\kappa$ (undistinguishable, within errors for $N=4,8,16$).
Large errors at $h =$ 0.7 and 0.8
are due to the sign problem at $h<1$.  }
 \label{susc} 
\end{center}
\end{figure}

On Figure\cite{Giedt:2007qg} \ref{susc}, we show the behavior of the  scalar field
  susceptibility as a function of $h$. The susceptibility, which can also be thought of as the zero-momentum propagator of $\phi_x$ (or the square of the correlation length), is defined in the usual way:
\beq
\label{chiscalar}
\chi = \frac{1}{N^2} \bigvev{ \left| \sum_x \; \phi_x \; \right|^2 }~,
\eeq
where the brackets indicate an expectation value computed with the partition function (\ref{mirrorZ1}), on a square $N\times N$ lattice.
As the figure shows, 
the correlation length is of order the lattice spacing for $h>1$. The study of (\ref{chiscalar}) as well as of other susceptibilities (composed from fermion bilinears) in ref.\cite{Giedt:2007qg} show that the theory is in the strong-coupling symmetric phase for $h>1$ and small $\kappa$ (at large $\kappa$, on the other hand, the theory is in the ``broken" phase).

\subsubsection{The  polarization operator and transversality}
 \label{transversality10}

The spectrum of theories with GW fermions can be  difficult
to determine, particularly when the theory is in a strong-coupling phase. 
Numerical simulations are often restricted to states created by particular operators and
do not lead to an exhaustive search for all possible light
degrees of freedom that may  exist. For example, in\cite{Giedt:2007qg} a search for massless mirror states
charged under the $U(1)$ (would-be-gauge) symmetry, created by several classes of operators, yielded a negative result. This created a puzzle, since in the absence of such states, the long distance physics of the ``1-0" model would then be described by an anomalous unbroken gauge theory---a situation argued to be impossible without a violation of some key principle (e.g., unitarity).

Clearly, in studies of mirror-fermion decoupling, 
  we are particularly  interested in states charged under the gauge symmetry\footnote{For   anomaly matching arguments for global symmetries that will not be gauged, as in Section \ref{summary}, we  imagine first weakly gauging the relevant global chiral symmetry, finding the relevant polarization operator  and then turning off the gauge field, as in (\ref{p1}).} (here: $U(1)$). The photon vacuum polarization operator, defined as:
 \begin{equation}
\label{p1}
\Pi_{\mu \nu}(x,y) \equiv {\delta^2 \ln Z[A] \over \delta A_\mu(x) 
	\delta A_\nu(y) }\bigg\vert_{A=0}~,
\end{equation}
can serve as a  probe for all possible light degrees of freedom
that are coupled to the gauge field. Below, we first study properties of the polarization operator of any theory with a gauge invariant partition function, i.e:
\begin{equation}
\label{z1}
\ln Z[A + \delta_\omega A] = \ln Z[A]~,
\end{equation}
with  $\delta_\omega A_\mu(x) = - \nabla_\mu \omega_x$, 
and $\nabla_\mu \omega_x = \omega_{x+\mu} - \omega_x$. This implies that:
\begin{equation}
\label{z2}
\sum_{\mu x} \frac{\delta \ln Z[A]}{\delta A_\mu(x)}\;  
\nabla_\mu \omega(x) = 0~.
\end{equation}
Taking ${\delta \over \delta \omega(x)}$ of (\ref{z2}) gives:
\begin{equation}
\label{z3}
\sum\limits_{\mu} \nabla^*_{\mu x} \; \frac{\delta \ln Z[A]}{\delta A_\mu (x)} = 0~,
\end{equation}
which, by expanding in $A_\mu$ around $A_\mu = 0$, implies transversality of 
all $n$-point functions:
\begin{equation}
\label{z4}
\sum\limits_\mu \nabla^*_{\mu x} \frac{ \delta^n Z[A]}{\delta A_\mu (x) \delta A_{\mu_1}(x_1) \ldots 
\delta A_{\mu_{n-1}} (x_{n-1})} \bigg\vert_{A=0}= 0~.
\end{equation}

For future use, note that, more generally,  if $O$ and $F$ are any functions of the gauge field $A$, not
necessarily gauge invariant, under the gauge transformation $\delta_\omega$,
$A\rightarrow A-\nabla_\mu \omega$, 
\begin{equation}
\delta_\omega O(A)=\sum\limits_\mu \frac{O(A)}{\delta A_\mu(x)}\nabla_\mu \omega(x)\,.
\end{equation}
Therefore, the following identity:
\begin{equation}
\label{eq:divergence}
\sum\limits_\mu \nabla_\mu ^\ast\left( F\cdot\frac{\delta O}{\delta A_\mu(x)}\right)
=\frac{\delta}{\delta\omega(x)} F \cdot \delta_\omega O
\end{equation}
is always true assuming only that $O$ and $F$ are smooth functions
of the gauge field.

The transversality condition (\ref{z4}) is only valid when it is applied to the full partition 
function of a theory where the gauge symmetry is respected, such as the
``1-0" model  or any other vectorlike theory (we have 
assumed that $Z[A]$ is 
a smooth function of $A$ at least in the vicinity of $A=0$; this
is always a valid assumption in vectorlike theories
provided that the action is smooth with respect to the gauge background
since the fermion measure is well defined).

\subsubsection{Further properties and exact relations obeyed  by the mirror $\Pi_{\mu\nu}$}
\label{mirrorpimunu}

Here, we present what will turn out to be (see the next Section) a lengthy derivation of the anomaly matching condition
on the lattice for the simple ``1-0" model. In the end of this Section, we also list some exact  (i.e., independent of couplings) properties of the mirror polarization operator.

The anomaly matching condition follows  from the transversality condition (\ref{z4}) of the vectorlike theory and the possibility to (locally) smoothly split the partition function into light and mirror parts. 
Thus, as  in (\ref{z01}), we have:
\begin{equation}
\label{z6}
\ln Z[A] = \ln Z_{light}[A] - \ln J[A] + \ln Z_{mirror}[A]~,
\end{equation}
where the Jacobian of the change of variables (\ref{psitoc}) is given\cite{Poppitz:2007tu}, in the charged sector of the ``1-0" model $(\psi_{\pm})$, by: 
\beq
J[A] =  {\rm det} || w_i(x)^\dagger u_j(x)^\dagger ||\;{\rm det} || v_i (x) t_j  (x) ||~.
\eeq
Note that $|| v_i(x) t_j(x) ||$ is a $2N^2 \times 2N^2$ dimensional matrix, with $x$ indexing rows and $i,j$-columns and that there is a similar, but $A$-independent, Jacobian  arising in the neutral sector ($\chi_{\pm}$). From (\ref{z6}), it follows that the polarization operator (\ref{p1}) also splits  
into ``light" and ``mirror'' parts:
\begin{equation}
\label{z61}
\Pi_{\mu \nu}(x,y) =\Pi_{\mu \nu}^{light}(x,y) + \Pi_{\mu \nu}^{mirror}(x,y)~.
\end{equation}
We include the Jacobian contribution into the ``light'' sector 
$\Pi_{\mu\nu}^{light}$ as a convention.

Now, under an arbitrary infinitesimal change $\delta_\eta$ of the gauge 
background, the ``light'' partition function 
$Z_{light}[A] = \det (u_i^\dagger[A] \cdot D[A] \cdot v_j)$ transforms (see\cite{Poppitz:2007tu} for a detailed derivation) as:
\begin{eqnarray}
\label{z7}
\delta_\eta \ln Z_{light}[A]= {\rm Tr}( P_+ D^{-1} \delta_\eta D) 
+ \sum_j (\delta_\eta u_j^\dagger \cdot u_j)  = {\rm Tr} ( P_+ D^{-1} \delta_\eta D) + j_\eta^u[A],
\end{eqnarray} 
where the currents $j_\eta^{u, w}[A]$ are  defined as:
\begin{equation}
\label{measurecurrent}
j_\eta^u[A] \equiv \sum_j (\delta_\eta u_j^\dagger \cdot u_j) ~, ~~ 
j_\eta^w[A] \equiv \sum_j (\delta_\eta w_j^\dagger \cdot w_j) ~.
\end{equation}
In terms of the currents (\ref{measurecurrent}), 
  the variation of the Jacobian is:\cite{Poppitz:2007tu}
\begin{equation}
\label{z8}
\delta_\eta \ln J[A] =   j_\eta^w[A] + j_\eta^u[A] ~.
\end{equation}
By combining (\ref{z7}) and (\ref{z8}), we find that the ``light'' plus 
Jacobian contribution to the change  of (\ref{z6}) under a 
gauge transformation is:
\begin{equation}
\label{z9}
\delta_\omega \ln {Z_{light}[A]\over J[A]} =- j_\omega^w[A] - {i \over 2} 
\sum_x \omega_x {\rm tr} \hat\gamma_{5 \; xx}[A]~,
\end{equation}
where $j_\omega$ denotes the measure current (\ref{measurecurrent}), restricted to a gauge variation of the background.
Now we take $\delta\over \delta \omega_x$ of (\ref{z9}) 
and use the identity \eqref{eq:divergence} to find:
\begin{eqnarray}
\label{z11}
 \sum\limits_\mu \nabla^*_{\mu x} \; 
{\delta \ln (Z_{light}[A] J^{-1}[A]) \over \delta A_\mu (x)} &=& 
- \sum\limits_\mu \nabla^*_{\mu x} \sum_i ( \delta_{\mu x} w_i^\dagger \cdot w_i) 
- {i \over 2} {\rm tr} \hat\gamma_{5 xx}[A] ~,\nonumber \\
 &=& - \sum\limits_\mu \nabla^*_{\mu } j_\mu^w[A]  
- {i \over 2} {\rm tr} \hat\gamma_{5 xx}[A] ~.
\end{eqnarray}
We have introduced the shorthand notation that we will
use freely hereafter:
 \begin{equation}
 \label{measurecurrent1}
 \delta_\mu \equiv {\delta   \over \delta A_\mu(x)}, 
 \end{equation}
where we suppressed the space-time index and understand that 
it is included in a single symbol $\mu$, as long as no ambiguities arise.  
Expanding (\ref{z11}) around $A_\mu = 0$ to linear order leads to:
\begin{equation}
\label{z13}
\sum\limits_\mu \nabla^*_{\mu x} \Pi^{light}_{\mu \nu}(x,y) 
= - \sum\limits_\mu \nabla^*_\mu \; \delta_\nu  j_\mu^w[A]
\bigg\vert_{A = 0} - {i \over 2} {\delta_\nu {\rm tr} \; 
	\hat{\gamma}_{5 xx}[A] }\bigg\vert_{A = 0}~,
\end{equation}
showing that the ``light'' polarization operator is not transversal. 
There are two contributions to its divergence: the first term, proportional 
to  the derivative of the ``measure current,"   exactly cancels with the 
identical contribution of the ``mirror" sector, as will become clearer later.

The second term on the r.h.s. of (\ref{z13}), proportional to the 
derivative of the topological lattice field tr$\hat\gamma_{5 xx}$ (that this is a topological lattice field follows from eqn.~(\ref{index}), which shows that $\sum_x$ tr$\hat\gamma_{5 xx}$ is an integer and thus can not change under a small variation of the gauge background)
represents the anomaly of the ``light'' fermions. To make contact with the anomaly in the continuum, we note 
that the topological lattice field can be expressed  
as:
\begin{equation}
\label{tl1}
{\rm tr}\;  \hat\gamma_{5 xx} = 
- {1 \over  2 \pi} \epsilon_{\mu\nu} F^{\mu \nu} + \nabla^*_\mu h^\mu[A]~,
\end{equation}
where $F_{\mu\nu} =\nabla_\mu A_\nu(x) - \nabla_\nu A_\mu(x)$ 
is the field strength $A_\mu$ and
$h^\mu[A]$ is a gauge invariant local current, and $\epsilon_{12}=1$. 
An explicit form of $h^\mu[A]$ can be obtained with some work.

By the local smoothness of the ``light"--``mirror" split, the fact that the full partition function of 1-0 model
is gauge invariant implies the  divergence  of the 
``light'' and ``mirror'' polarization operator cancel exactly.
Hence, from (\ref{z11}), we conclude that 
\begin{equation}
\label{mirrorPdivergence}
\sum\limits_\mu \nabla^*_{\mu x} \Pi^{mirror}_{\mu \nu}(x,y) 
= \sum\limits_\mu \nabla^*_\mu \; \delta_\nu  j_\mu^w + \frac{i}{2} \delta_\nu {\rm tr} \; 
	\hat{\gamma}_{5 xx}~.
\end{equation}
Since the terms that depend on the measure current in the ``light'' 
and ``mirror'' polarization operators cancel, they can be
considered a lattice artifact due to the choice of the
fermion measure.  The last term in the above equation is more 
important since it can be expressed in terms of some 
correlation functions in
the mirror sector and therefore represents physically meaningful quantities.
The fact that its divergence should agree exactly with that
of the ``light'' sector appears to suggest that
there always exist a certain light degree of freedom, even 
in the mirror sector, in order to contribute the appropriate
anomalous divergence of $\Pi^{mirror}_{\mu\nu}$. 

We note that the conclusion that the mirror polarization operator is non-transverse, and that its divergence is proportional to the variation of the topological lattice field, 
 depends only on the fermion content
in the ``mirror'' sector and not on details on the mirror-theory dynamics, such as the strength of its couplings. 
We will argue in Section \ref{matching} that eqn.~(\ref{mirrorPdivergence}) is the lattice version of the
t' Hooft anomaly matching condition in the GW formalism in 2d. 

Equations similar to (\ref{z13}, \ref{mirrorPdivergence}) also hold  for 
the non-transverse higher derivatives of the light and mirror partition 
functions. However, in 2d,   to study the interplay 
between the anomaly and the light degrees of freedom it is sufficient to  consider  only the polarization operator in a 
trivial gauge background. 
In 4d, the trivial gauge background analysis of this Section would have to be 
extended to the three-point function in order to capture the 
effect of the anomaly. The mirror polarization operator would still be useful to detect massless mirror states in 4d.

To end this Section, we now list some exact properties of polarization operators that hold for general chiral theories, in particular for our mirror theory; these are derived in Appendix A of ref.\cite{Poppitz:2009gt} 
All equations below refer to the polarization operator in $x$-space and not to their Fourier transforms; recall that we absorb the space-time indices into $\mu$, $\nu$ (see (\ref{measurecurrent1})).
The definition of  $\Pi_{\mu\nu}^{mirror} = \delta_\mu \delta_\nu \log Z_{mirror}$, see (\ref{z61}, \ref{z6}),  implies that $\Pi_{\mu\nu}^{mirror}$ is symmetric due to local smoothness of $Z_{mirror}$. 

For the discussion here, the most important property is that the mirror polarization operator $\Pi_{\mu\nu}^{mirror}$ can be decomposed into a part that is a measure current derivative and the rest:
\beq
\label{pgeneral1}
\Pi_{\mu\nu}^{mirror} = \delta_\nu j_\mu^w + \Pi_{\mu\nu}^{mirror \; \prime}~.
\eeq
As shown in\cite{Poppitz:2009gt}, $\Pi_{\mu\nu}^{mirror \; \prime}$ is always a total derivative:
\beq
\label{pgeneral2}
\Pi_{\mu\nu}^{mirror \; \prime} = \delta_\nu \Pi^{mirror \; \prime}_\mu~.
\eeq
and, in addition, $\Pi_\mu^{mirror \; \prime}$ is exactly gauge invariant.
Therefore, proceeding as in the derivation of (\ref{z11}), we find:
\beq
\label{pgeneral3}
\nabla^*_\nu \Pi_{\mu \nu}^{mirror \; \prime} = 0~,
\eeq
while with respect to the first index, we have:
\begin{equation}
\label{abc2}
\nabla^\ast_\mu \Pi^{mirror \; \prime}_{\mu\nu}=\frac{i}{2}\delta_\nu \tr \hat \gamma^5_{xx}.
\end{equation}
Now, the total mirror $\Pi_{\mu\nu}^{mirror}$ is symmetric, but $\delta_\nu j_\mu^w$ and $\Pi_{\mu\nu}^{mirror \; \prime}$ are separately not, but obey:
\beq
\label{pgeneral4}
  \left( \Pi^{mirror \; \prime}_ {\mu \nu} -\Pi^{mirror \; \prime}_ {\nu \mu} \right) = - \delta_\nu j_\mu^w + \delta_\mu j_\nu^w = {\cal{F}}_{\mu \nu}~,
\eeq
where ${\cal{F}}_{\mu \nu}$ is the curvature of the measure current, which is a known local functional of the gauge field whose divergence $\nabla_\mu^* {\cal{F}}_{\mu \nu}$ gives half the anomaly. 
These results imply that the symmetric and antisymmetric parts of $\Pi^{mirror \; \prime}_{\mu \nu}$ each contribute half of the anomalous divergence (\ref{abc2}). Since all properties listed in this Section hold independently of the mirror action, in particular of the strength of the mirror couplings, the verification of  (\ref{pgeneral3}), (\ref{abc2}), and (\ref{pgeneral4}), in a numerical simulation at strong mirror couplings provides an important check on its consistency.

Finally, we stress that in our numerical simulations, we calculate $\Pi_{\mu\nu}^{mirror \; \prime}$, as the measure-current part of 
$\Pi_{\mu\nu}^{mirror}$, see (\ref{pgeneral1}), is exactly the opposite that of the light theory. Obtaining an expression for  $\Pi_{\mu\nu}^{mirror \; \prime}$ in terms of mirror-theory correlation functions, which can be computed using Monte Carlo methods, is a somewhat arduous task which is accomplished by repeated use of the splitting theorem (see Section \ref{splitZ}); the calculation is explained in detail in\cite{Poppitz:2009gt}.
 
\subsection{Anomaly matching and  its possible solutions}
\label{matching}

As already explained, in what follows, we focus on the  $\Pi^{mirror \; \prime}_{\mu\nu}$ contribution to  the mirror theory polarization operator. 
In fact, we will focus only on
its symmetric part, since its antisymmetric part, as we discussed in Section \ref{mirrorpimunu}, is of little physical interest---as it is independent on details of the interactions and depends only on the field content, see eqn.~(\ref{pgeneral4}), it contains no information about the mirror spectrum.

 Here, we will study both the real and imaginary parts of the 
polarization operator of the mirror sector. We will show that  
its imaginary part has a nonlocal contribution that gives rise to the anomaly. 
  The real part of the polarization operator is a universal probe of  
the number and nature of the massless charged degrees of freedom 
and thus gives information of the spectrum. 

Let us first explain the nonlocality of the imaginary part of the polarization operator.
For brevity, below we omit the superscripts (prime and mirror) and simply
refer to $\Pi^{mirror \; \prime}_{\mu\nu}$  as $\Pi_{\mu\nu}$. We work in  momentum space and denote Fourier transformation
of $\Pi_{\mu\nu}$ as $\tilde \Pi_{\mu\nu}$.
Let us first explore the consequence of equation (\ref{mirrorPdivergence}), which, together with (\ref{pgeneral1}) and (\ref{tl1}),  implies that for small momentum $q$: 
\begin{equation}
\label{piminus1}
i q^\mu \tilde\Pi_{\mu \nu}(q) 
	= {1 \over 2 \pi} \epsilon_{\nu \lambda} q^\lambda+~ {\cal{O}}(q^2)~.
 \end{equation}
The rhs of (\ref{piminus1}) is of course local.
One wonders if the usual argument in the continuum that it can not be 
the divergence of a  local expression applies on the  lattice too. 
A quick argument showing that it does is as follows. 
Let $c=-i/(2\pi)$. Ignoring the real and transversal part 
of $\tilde \Pi_{\mu\nu}$, we can rewrite (\ref{piminus1}) as the set of 
two equations:
 \begin{eqnarray}
 \label{piminus2}
 \tilde\Pi_{22}  &=& - (c  + \tilde\Pi_{12}^-)\; {q_1 \over q_2} ~,\\
 \tilde\Pi_{11} &=& (c - \tilde\Pi_{21}^-) \;{q_2 \over q_1} 
 = (c - \tilde\Pi_{12}^-) \;{q_2 \over q_1}, \nonumber
  \end{eqnarray}
where we used $\tilde\Pi_{12} =\tilde\Pi_{21}$.
Locality of $\Pi_{11}$ and $\Pi_{22}$ 
 would require that $c-\tilde{\Pi}_{12}  = A q_1 + ...$, $c+ 
 \tilde{\Pi}_{12}  = B q_2 + ...$, where $A$ and $B$ are arbitrary 
 constants and dots denote higher powers of momenta. This 
 leads to:
 \begin{eqnarray}
\label{piminus3}
\tilde\Pi_{12}  &=& -c + B q_2 + {\cal{O}} (q^2)~,\\
 \tilde\Pi_{12}  &=& c - A q_1 +{\cal{O}} (q^2)~,\nonumber
\end{eqnarray}
conditions, which are clearly incompatible. 

On the other hand, a nonlocal solution of the anomaly 
conditions (\ref{piminus2}) is given by 
$\tilde\Pi_{12}  = c (q_2^2 - q_1^2)/(q_1^2 + q_2^2)$. 
This is, indeed, the form of the non-transverse part of 
the polarization operator for an anomalous theory in the continuum, where
the general solution of $\Pi_{\mu\nu}$ in 2d is  given by:
\begin{equation}
\label{pigeneral}
\Pi_{\mu\nu}(q)=C_1(q^2)\left(\frac{q^\mu q^\nu}{q^2}-\delta^{\mu\nu}\right)
	+C_2(q^2)\frac{\epsilon^{\nu\rho}q_\rho q^\mu+\epsilon^{\mu\rho} 
		q_\rho q^\nu}{q^2},
\end{equation}
where $C_1$ and $C_2$ are two arbitrary functions.

We now enumerate the possible forms for $C_1$ and $C_2$.
As mentioned above, in Euclidean space the anomaly appears in the 
imaginary part of the polarization operator only.
It is well-known in the continuum that in a unitary 
Lorentz (Euclidean) invariant theory the zero-momentum singularity 
in the solution of (\ref{piminus1}) is due to either a massless 
Goldstone boson or a massless fermion\cite{Frishman:1980dq,Coleman:1982yg}. 
The massless scalar or fermion will, of course, also give a nonlocal,
but divergence free, contribution to the real part of the polarization operator.
If the GW formalism is to lead to any reasonable continuum limit, we would expect
this conclusion to remain true on lattice.

In order to see what nonlocal contributions to the real part of the polarization operator to expect, consider first a ``Green-Schwarz" scalar theory in 2d
Euclidean space:
\begin{equation}
\label{scalar1}
Z_{GS}[A] = \int {\cal{D}} \eta \; e^{\int d^2 x (-{\kappa\over 2} 
		(\partial_\mu \eta - A_\mu)^2 + i { \eta\over 2 \pi} F_{12}) }~.
\end{equation}
Explicit computation leads to:
\begin{equation}
\label{scalar3}
\tilde\Pi^{\mu\nu}_{GS}\big\vert_{A = 0}(q) 
	=\left(\kappa + {1 \over 4 \pi^2 \kappa}\right)  
	\left( {q^\mu q^\nu \over q^2} - \delta^{\mu\nu} \right)
	-  {i \over 2 \pi} \;
{\epsilon^{\nu \rho} q_\rho q^\mu + \epsilon^{\mu \rho} q_\rho q^\nu \over q^2}~.
\end{equation} 
The polarization operator is not transverse and its divergence is 
$ i q_\mu \tilde\Pi^{\mu\nu}_{GS}\big\vert_{A = 0}(q) 
	= {1\over 2 \pi}  \epsilon^{\nu \rho} q_\rho$, identical to
(\ref{piminus1}). 
The real part of (\ref{scalar3}) contains obviously the contribution of 
a massless scalar; notice the shift of the coefficient 
$\kappa \rightarrow \kappa + {1 \over 4 \pi^2 \kappa}$ 
due to the anomalous Green-Schwarz term. If our mirror 
theory has a massless Green-Schwarz scalar, it would have to 
manifest itself by contributing to both the real and imaginary 
parts of the mirror polarization operator in agreement with (\ref{scalar3})
in the small momentum limit.

Similarly, a massless mirror charged chiral fermion would 
also  contribute to both the real and imaginary parts of $\Pi_{\mu\nu}$:
\begin{equation}
\label{fermion}
\tilde\Pi^{\mu\nu}_{chiral}\big\vert_{A = 0}(q) 
= {1 \over 2  \pi} \left( {q^\mu q^\nu \over q^2} - \delta^{\mu\nu} \right)
-  {i \over 2 \pi} \;
{\epsilon^{\nu \rho} q_\rho q^\mu + \epsilon^{\mu \rho} q_\rho q^\nu \over q^2}~,
\end{equation} 
where the   real part of $\tilde\Pi_{\mu\nu}(q)$  of the chiral fermion 
is  equal to one-half that of the Dirac fermion in the Schwinger model.
 Eqn.~(\ref{fermion}) is the small-momentum limit of the contribution of a  free chiral 
GW fermion  to the basis-vector independent part of the polarization 
operator, i.e., exactly what a massless mirror at $y=0$ 
would contribute to the full polarization operator of the vectorlike theory; see\cite{Poppitz:2009gt}. 
Note that, as opposed to  the Green-Schwarz scalar (\ref{scalar3}), 
the coefficients of the real and imaginary parts of the massless 
chiral fermion $\tilde\Pi_{\mu\nu}$ are  the same.\footnote{We should note that in {\it continuum} 2d, the two realizations of anomaly matching (\ref{fermion}) and (\ref{scalar3}) are equivalent by bosonization 
(see\cite{Halliday:1985tg} for the study of anomaly free 2d ``Pythagorean" models as well as the more recent ref.\cite{Kutasov:1994xq}). Our findings of Section \ref{symmetryvsexplicit} of massless fermions vs  massless scalars in the strongly-coupled mirror sector of the ``1-0" model should be really phrased as ``in the lattice formulation that we use the results are naturally described as due to either a `goldstone' or a chiral fermion." If we knew how to bosonize a lattice fermion with a Neuberger-Dirac operator, perhaps the equivalence could be extended to finite lattice spacing.}

If long-distance unitarity was violated, one could imagine that the imaginary part of the polarization operator had a $1\over q^2$ pole, providing the correct anomalous divergence, but the real part did not---as if there was no particle in the spectrum responsible to the anomaly.
Our Monte Carlo simulations for the ``mirror'' sector
did not find any evidence of such kind. Instead, it was found
that  unitarity  is respected anywhere in
parameter space, so that any chiral lattice theory
with GW fermions (e.g., one with strong Yukawa interactions) appears always to lead to a physically reasonable field theory
in the continuum limit.

   \subsection{The symmetry arguments vs explicit lattice simulations  }
\label{symmetryvsexplicit}

Before we describe the results of the  Monte Carlo simulation of the mirror sector of the ``1-0" model, let us briefly explain the idea how this was
done.
 
In the continuum,  the contribution to the Fourier transform of the real 
part of the polarization operator due to massless particles takes the
form (\ref{pigeneral}):
\begin{equation}
\label{picont1}
\tilde\Pi_{\mu\nu}(k) = 2 C
\;   {\delta_{\mu\nu} k^2 - k_\mu k_\nu\over k^2}
\end{equation}
where we have rescaled $C_1=2C$ for convenience and denoted the momentum by $k$. For a single
massless degree of freedom, we have from (\ref{scalar3}):
\begin{equation}
\label{scalardisc}
2 C_{GS\; scalar} \simeq - {\kappa}
\end{equation}
for a Green--Schwarz scalar, and for a single charged Weyl fermion from (\ref{fermion}):
\begin{equation}
\label{fermiondisc}
2C_{ch.ferm.} \simeq-{1 \over 2 \pi} \simeq -0.159~.
\end{equation}
The constant $C$ in (\ref{picont1}) thus reflects the number and nature of massless degrees 
of freedom in the theory.  To help identify the 
form of $\tilde \Pi_{\mu\nu}$ and the coefficient
$C$, we notice that the polarization operator (\ref{picont1})    
in 2d has a directional singularity as $k \rightarrow 0$ because:
\begin{eqnarray}
\label{picont2}
\tilde\Pi_{11}(\phi)\big\vert_{k \rightarrow 0} &=&  
C (1 - \cos 2 \phi)~ \nonumber \\
\tilde\Pi_{21}(\phi)\big\vert_{k \rightarrow 0} &=&  -C  \sin 2 \phi ~, 
\end{eqnarray}
where $\phi$ is the polar angle in the $k_{1,2}$ plane, 
so that $\tilde \Pi_{\mu\nu}$ approaches a different limit
as $k$ approaches $0$ from different directions.
It is this angular singularity that can be easily identified
in lattice simulations.  In comparison, if only massive particles
exist in the spectrum, the low momentum limit of $\tilde \Pi_{\mu\nu}$
would be proportional to $(k^\mu k^\nu-k^2 g^{\mu\nu})/m^2$ and
the angular singularity is absent.

\begin{figure}
\begin{center}
\includegraphics[width=5in]{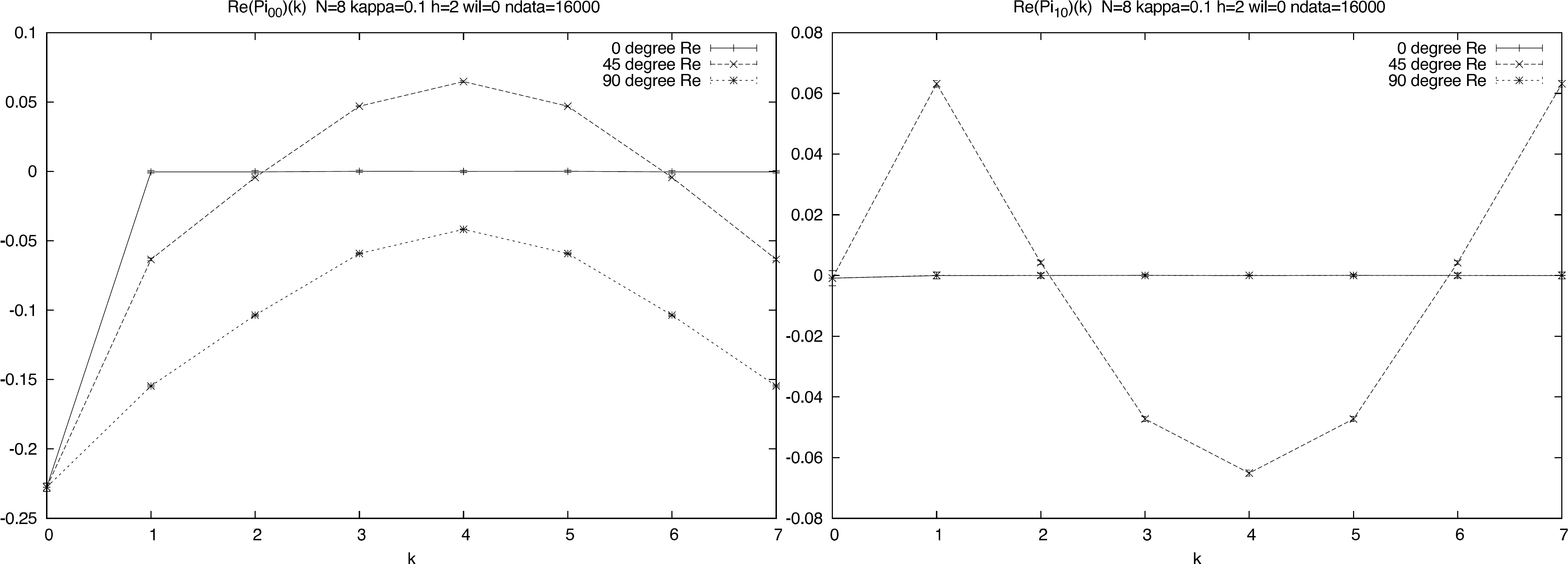}
\caption
{The real parts of $\Pi_{00}$ and $\Pi_{10}$ of the mirror 
for $\kappa = 0.1$, $h=2$, as a function of momentum approaching 
the origin in  different directions. The value of the discontinuity indicates that when the system is in the strong-coupling 
symmetric phase, a single massless charged chiral  fermion exists.
\label{p_0.1_2} }
\end{center}
\end{figure}

\begin{figure}
\begin{center}
\includegraphics[width=5in]{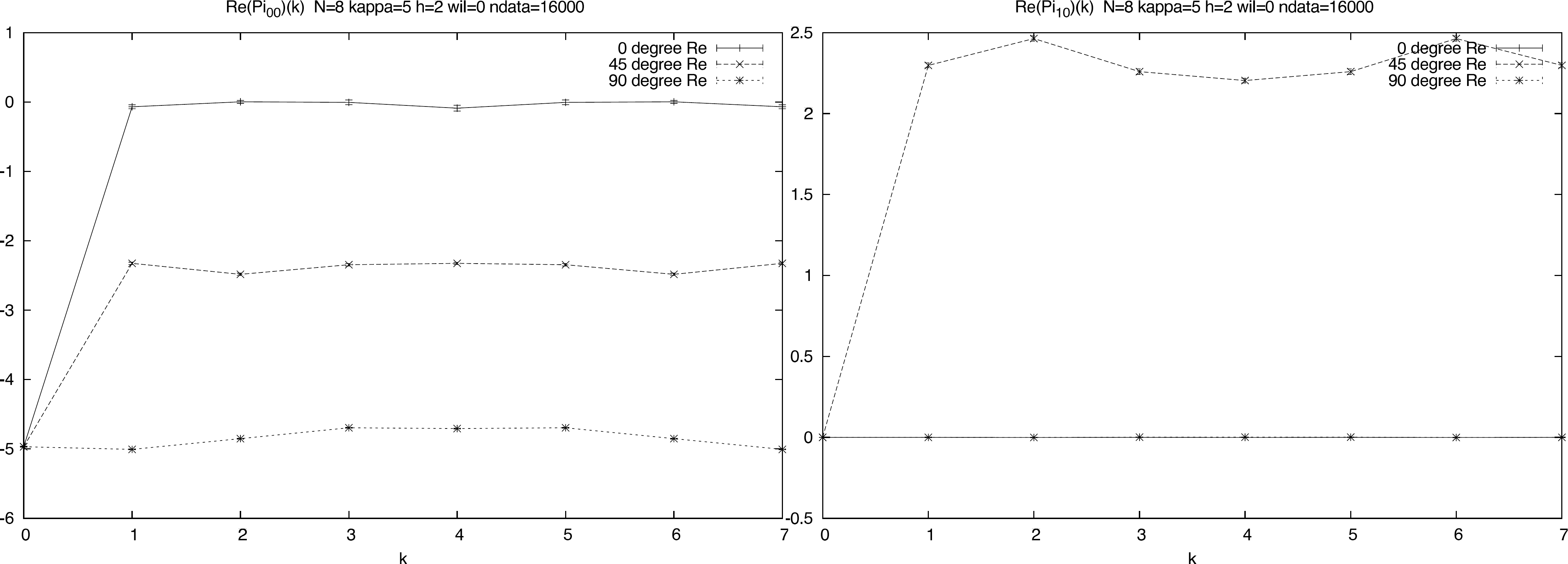}
\caption
{The real parts of $\Pi_{00}$ and $\Pi_{10}$ of the mirror for 
$\kappa = 5$, $h=2$ indicating the system is in a ``broken" (algebraically ordered) phase where
a single massless ``goldstone" scalar appears, as explained in\protect\cite{Poppitz:2009gt}.
}
 \label{p_5_2} 
\end{center}
\end{figure}

Monte-Carlo simulations have been done for the mirror sector
of the ``1-0" model and the behavior the photon vacuum
polarization operator in this theory has been investigated.
The detailed results are presented
in\cite{Poppitz:2009gt}.  It suffices to mention here that
the system appears to adjust itself ``automatically" in different
regions of the parameter space, such that a single massless
degree of freedom is always present, responsible for satisfying
the anomaly equation. The angular singularity and the value of the discontinuity $C$  (\ref{picont2}) of the real
part of $\tilde \Pi_{\mu\nu}$ that was found suggests that  unitarity  was respected. 
Furthermore, the value of $C$ was found to agree\footnote{The reader can consult ref.\cite{Poppitz:2009gt}, or, alternatively, infer the small-$k$  discontinuities (\ref{picont2}) at $\phi$$=$$0^o$, $45^o$, $ 90^o$ from   the plots on Figs.~\ref{p_0.1_2} and \ref{p_5_2} (replacing the index $0$ by $2$) and then compare with the corresponding continuum values (\ref{scalardisc}, \ref{fermiondisc}).} well with either (\ref{scalardisc}) or (\ref{fermiondisc}), depending on the phase of the mirror theory, allowing us to interpret the result as due to either a ``goldstone" boson or a chiral fermion. The agreement with the continuum values was impressive, given the rather small $8^2$ lattice used in the simulation (the complexity of the calculation of $\Pi_{\mu\nu}$ restricted us to rather small volumes). 

When the system is in the strong-coupling symmetric phase, where no massless 
scalar is expected, the result showed the consistency with
the prediction of a single massless chiral GW fermion as
shown in Figure~\ref{p_0.1_2}. In this Figure, the behavior
of $\tilde \Pi_{\mu\nu}$ with $k$ approaching  zero in three
different directions  is shown. The entire plot can be compared to the prediction
of a single free GW chiral fermion, whose contribution to the
divergence of $\tilde \Pi_{\mu\nu}$
can be analytically calculated. It was found\cite{Poppitz:2009gt} that the two
discontinuities agree very well. 

When the system is
in the ``broken" phase (at large $\kappa$), where a single massless scalar should
appear, the simulation also found the result to be consistent
with the prediction given by a single Green-Schwarz scalar, 
as shown on Figure~\ref{p_5_2}.

Furthermore, when $h \rightarrow 0$ in the symmetric phase, the discontinuity  (\ref{picont2}) found in the Monte Carlo simulations was found to be approximately that appropriate for three massless chiral fermions. These are a massless charged chiral fermion and a massless charged vectorlike pair, whose appearance at $h=0$ can be analytically understood and will be explained in Section \ref{domainwall}. 

To summarize, the numerical evidence available so far suggests that strong Yukawa models formulated with GW fermions respect 't Hooft anomaly matching. Most importantly, in the strong coupling symmetric phase, where all mirror symmetries (but the one to be gauged) are explicitly broken,  the minimal number of charged fermions required to saturate the 't Hooft anomaly of the to-be-gauged $U(1)$ remains massless. One can view the results of this Section as a numerical ``proof" of 't Hooft anomaly  matching and as evidence for the long-distance unitarity of GW-fermion Yukawa models.

  \subsection{A different form of the split partition function; the relation to domain wall fermions}
 \label{domainwall}

In this Section, we will attempt to get a  better analytic understanding of some of the numerical results of the previous Sections. To this end, we will develop a different representation of the 1-0 model partition function and comment on its relation to domain-wall fermions. 
A representation like the one considered here, as well as the comments we make, hold  for any GW-fermion model attempting to decouple mirror fermions from a vectorlike gauge theory by means of some non-gauge strong interactions.

We begin by noting that the ``1-0" model action (\ref{10modelaction})  can be written in unprojected components (the superscripts $\hat{P}^{1 (0)}$ in the hatted projectors indicate whether the charge-1 or charge-0 Dirac operator has to be used):
\begin{eqnarray}
\label{toymodel2}
S_{kinetic}&=& - \left( \bar\psi   D_1     \psi \right) - \left( \bar\chi    D_0    \chi \right) \nonumber  \\
 S_{Yukawa} &=&  y \left\{ \left( \bar\psi \hat{P}_+^1   \phi^*  P_+  \chi  \right) + \left( \bar\chi \hat{P}_-^0    \phi {P}_- \psi  \right) \right\}  \\
  &+& y h \left\{ \left( \psi^T (P_-)^T  \phi \gamma_2   P_+  \chi \right) - \left( \bar\chi \hat{P}_-^0  \gamma_2  \phi^* (\hat{P}_+^1)^T  \bar\psi^T \right)  \right\} \nonumber~,
\end{eqnarray}
where the measure now is the usual vectorlike theory measure in terms of $\psi, \bar\psi, \chi, \bar\chi$:
\begin{eqnarray}
\label{Ztotal}
Z = \int d \psi d \bar\psi d \chi d\bar\chi d \phi \; e^{- S_{kinetic} - S_{Yukawa} - S_\kappa} ~,
\end{eqnarray}
and $S_\kappa$ is defined in (\ref{Skappa}). 
We now perform the field redefinition:
\begin{eqnarray}
\label{fieldredef1}
\psi  \rightarrow \psi, ~  \bar\psi \rightarrow \bar\psi\; {1 \over 2 - D_1}, ~
\chi \rightarrow \chi, ~  \bar\chi \rightarrow \bar\chi \;{1 \over 2 - D_0}.
\end{eqnarray}
The motivation for this redefinition can be traced back to the GW relation, which  implies 
$
{1 \over 2 - D}\; \hat{\gamma}_5 = \gamma_5 \;{1 \over 2 - D} ,
$
in other words, a $\hat{\gamma}_5$  action on the original $\bar\psi$ is transformed into the action of  $\gamma_5$ on the new fields in (\ref{fieldredef1}).

The price to pay for having an action of the lattice chiral symmetries generated by $\gamma_5$, exactly as in the continuum, is the nonlocality of the redefinition (\ref{fieldredef1}) and thus of the resulting action (see (\ref{Ztotal1}) below). 
However, since we  work  perturbatively  in the gauge field and at finite volume, we will  imagine throughout this Section that the singularity at $D=2$ of (\ref{fieldredef1}) is  avoided by turning on background Wilson lines. The effect of the field redefinition on the partition function is:
\begin{eqnarray}
\label{Ztotal1}
Z &=& \int d \psi d \bar\psi d \chi d\bar\chi d \phi \; {\rm det}(2 - D_1) \; {\rm det}(2 - D_0) \; e^{ -S_{kinetic}^\prime - S_{Yukawa}^\prime - S_\kappa}~,\nonumber \\
S^\prime_{kinetic} &=&  - \left( \bar\psi  \; {D_1 \over 2 - D_1}  \;  \psi \right) - \left( \bar\chi  \;  {D_0 \over 2 - D_0}   \; \chi \right)~,
\end{eqnarray}
\begin{eqnarray}
\label{Ztotal12}
&&S^\prime_{Yukawa}  =    {y\over 2}\left( \bar\psi    \phi^*  P_+  \chi  \right) + {y\over 2}\left( \bar\chi      \phi {P}_- \psi  \right) \\
+&&  yh \left[ \left( \psi^T    \phi \gamma_2   P_+  \chi \right) - {1 \over 4} \left( \bar\chi   P_-  \gamma_2  \phi^* \bar\psi^T \right)    - {1 \over 4} \left( \bar\chi P_- {D_0 \over 2 -D_0}  \gamma_2  \phi^* ( {D_1 \over 2 - D_1})^T P_+   \bar\psi^T \right) \right]    \nonumber.
\end{eqnarray}
Obtaining the transformed Yukawa couplings in $S_{Yukawa}^\prime$ requires repeated use of the GW relation  and the equivalent relation shown in the paragraph after eqn.~(\ref{fieldredef1}). Needless to say, similar redefinitions hold in four dimensions and can be straightforwardly performed if necessary. 

Several comments, concerning the representation (\ref{Ztotal1}, \ref{Ztotal12}) of the mirror partition function, are now in order. We hope that these comments are useful to clarify the relation between different formulations of exact lattice chirality:
\begin{enumerate}
\item
 The singularity of the action at  the position of the doublers, $D=2$, ensures that they have  infinite action and do not propagate at the classical level,  as first proposed by Rebbi\cite{Rebbi:1986ra}. The problem with\cite{Rebbi:1986ra}, pointed out in\cite{Bodwin:1987gi,Pelissetto:1987ad}, of the doublers contributing as ghosts to  the photon polarization operator at the quantum level is solved by the determinant prefactors, which exactly cancel the would-be ghost/doubler contributions.
 \item Another comment concerns the relation of (\ref{Ztotal1}), with $y = h =0$, to  domain wall fermions. In the 2d case, these propagate on a finite interval in three dimensions. The generating functional of Green's functions of the boundary chiral modes---in the case at hand,   one charged and one neutral---can be obtained by integrating out  the bulk fermions. This is  technically possible since the fermion action is bilinear and there is no gauge field propagation in the extra dimension (or any other non-uniformity except at the boundaries). When taking the chirally-symmetric limit of an infinite number of sites in the extra dimension,  a massive bulk Pauli-Villars field, antiperiodic in the extra dimension, has to be included in order to obtain a finite determinant. The generating functional of Green's functions for the boundary chiral modes can be represented as a partition function with source terms. The result is exactly (\ref{Ztotal1}), with $y = h =0$, and with source terms for $\psi$ and $\chi$ included. The determinant prefactors in (\ref{Ztotal1}) arise as a combination of the determinants of the bulk fermions and the Pauli-Villars fields. The derivation of these results   can be extracted  from the work of ref.\cite{Kikukawa:1999sy}.
\item It is less straightforward to relate the domain-wall fermion construction to  models where strong GW-fermion Yukawa or multifermion interactions are present. In this regard, recall 
 the ``pre-GW" proposal of\cite{Creutz:1996xc}, where strong multifermion interactions were added on one domain wall with the intent to decouple the mirrors from the Standard Model. We note also that the ``warped domain wall" idea of ref.\cite{Bhattacharya:2005xa}, whose 2d formulation was studied in\cite{Bhattacharya:2006dc}, also uses strong interactions in a domain-wall set up.  It would be interesting to more explicitly relate the finite-size extra dimensional version of these ideas   (with strong ``brane-localized" multifermion of Yukawa interactions) to the present (or similar) construction, since understanding the relation may yield both practical benefits and theoretical insight.
\item For vanishing Majorana coupling, $h=0$, the   Yukawa interaction in (\ref{Ztotal1}) is equivalent to the chirally invariant Yukawa coupling of ref.\cite{Luscher:1998pq}, see also\cite{Chiu:1998aa}. To see this, 
use $(\bar\psi {D \over 2 - D} \psi) = {1 \over 2} (\bar\psi D \psi) + {1 \over 2} (\bar\psi D {D \over 2 - D} \psi)$  to replace the $\psi$ kinetic term in the action $S_{kin}^\prime$. Then note that $- {1 \over 2} (\bar\psi D {D \over 2 - D} \psi) = (\bar\xi (2 - D) \xi) + {1 \over \sqrt{2}} (\bar\xi D \psi) + {1 \over \sqrt{2}} (\bar\psi D \xi)$, where it is understood that the new charged field $\xi$ is integrated out from the action using its equation of motion, while the $\xi$-determinant exactly cancels  the one in (\ref{Ztotal1}). Then, shift the integration variable $\psi \rightarrow \psi + \sqrt{2} \xi$. Next, perform exactly the same operations on $\chi$, introducing a neutral field $\eta$, to finally obtain the  action in the original  form of L\" uscher, suitably adapted to the 2d case and to our normalization:
\beqa
\label{luscheryukawa} S &=& - {1\over 2}(\bar\psi D_1 \psi + \bar\chi D_1 \chi) + 
2 \bar\xi \xi + 2 \bar\eta \eta  \\
&+& {y \over 2}  (\bar\psi + \sqrt{2} \xi) \phi^* P_+ (\chi + \sqrt{2} \eta) +   {y \over 2}(\bar\chi + \sqrt{2} \eta) \phi P_- (\psi + \sqrt{2} \xi), \nonumber
\eeqa
where the measure is the trivial one over $\psi, \chi, \eta, \xi$.  The Yukawa interaction of (\ref{toymodel}) is thus equivalent to that of\cite{Luscher:1998pq}. We note that gauge and chiral invariant Majorana couplings were not considered   in\cite{Luscher:1998pq} and to the best of our 
knowledge were first constructed in\cite{Giedt:2007qg} (see also\cite{Igarashi:2009kj}).
\end{enumerate}

For our purposes, 
  the representation (\ref{Ztotal1},\ref{Ztotal12}) of the ``1-0" model partition function is useful to emphasize the importance of Majorana couplings. It will allow us to 
 understand the result mentioned at the end of Section \ref{symmetryvsexplicit} that at vanishing Majorana couplings, the solution of the 't Hooft anomaly matching condition is not minimal  and involves an extra vectorlike pair. The Majorana $h\ne0$ term gives mass to this extra massless vectorlike pair. 
 
We can split the partition function (\ref{Ztotal1},\ref{Ztotal12}) exactly as we did in the basis of GW fermions. The representation of the split partition function that we give below is, in fact, equivalent to that in (\ref{z01}).
  The splitting of the $\psi$-$\chi$ partition function (\ref{Ztotal1}) into a ``light" and ``mirror" part is now done trivially using the $\gamma_5$-eigenvectors, which have no gauge-field dependence. We obtain, denoting now by $\psi_\pm, \chi_\pm$ the ``normal" $\gamma_5$-chirality components of the 2-component $\psi,\chi$, and using the fact that $S^\prime_{Yukawa}$ only depends on the mirror components of $\psi, \chi$:
  \begin{eqnarray} \label{Ztotal2}
  Z &=& Z_+ \times Z_- \times {1 \over J} ~, \nonumber \\
  Z_+&=& {\rm det}_- (2 - D_0)  {\rm det}_+ (2 - D_1) \int d  \psi_+  d  \chi_-  e^{- \left( \bar\psi_+    {D_1 \over 2 - D_1}    \psi_+ \right) - \left( \bar\chi_-   {D_0 \over 2 - D_0}    \chi_- \right)  }  ~,\nonumber\end{eqnarray}
  where the mirror partition function\footnote{Here and below, we renamed $Z_{light} \equiv Z_{+}, Z_{mirror} \equiv Z_-$.} now is: 
  \beqa
  \label{Ztotal3}
  &&Z_-  =  \\
&& {\rm det}_+ (2 - D_0) {\rm det}_- (2 - D_1) \int d \psi_-  d  \chi_+ d \phi  e^{- \left( \bar\psi_-    {D_1 \over 2 - D_1}   \psi_- \right) - \left( \bar\chi_+   {D_0 \over 2 - D_0}    \chi_+ \right)  - S^\prime_{Yukawa} - S_\kappa}  ~.\nonumber
\eeqa
Splitting the determinant prefactor into ``light" and ``mirror," as indicated in (\ref{Ztotal2},\ref{Ztotal3}),   requires  using the gauge-field dependent eigenvectors of $\hat{\gamma}_5$:
\begin{eqnarray} \label{Ztotal4}
{\rm det}_+ (2 - D) &=&{\rm det}_+ (1 + \hat{\gamma}_5 \gamma_5)  =  {\rm det} || (u^\dagger_i (2-D)t_j)|| =  {\rm det} ||2 (u^\dagger_i t_j)|| , \\
{\rm det}_- (2 - D) &=& {\rm det}_- (1 + \hat{\gamma}_5 \gamma_5)=   {\rm det} ||  (w^\dagger_i (2 - D)v_j)||= {\rm det} ||2 (w^\dagger_i v_j)|| , \nonumber
\end{eqnarray}
 where the appropriate $\hat{\gamma}_5$-eigenvectors are to be used for $D^0$ or $D^1$.  
 
 The gauge variation of the mirror partition function $Z_-$ has now two contributions: one from the variation of ${\rm det}_- (2 - D^1)$  and one from the variation of the path integral over $\psi_-, \chi_+, \phi$. The splitting theorem, applied to the chiral partition function defined by the path integral in (\ref{Ztotal3})  can be easily seen to imply that the gauge variation   $Z_-$ with  the determinant factors left out vanishes when $h=0$ and $y \rightarrow \infty$. Thus, the entire gauge variation of $Z_-$ comes from the determinants:
\begin{eqnarray}
\label{vartn1}
\delta \ln Z_- &=& \delta_\omega {\rm ln det} ||  (w^\dagger_i (2-D^1)v_j)|| =  \sum_j (\delta_\omega w_j^\dagger  w_j ) + {\rm tr}\; \hat{P}_+ {1\over 2 - D^1} \delta_\omega(2- D^1) \nonumber \\
&=&   \sum_j (\delta_\omega  w_j^\dagger w_j ) + i  {\rm Tr}\omega \left(  \hat{P}_+^1 - P_+ \right) =  \sum_j ( \delta_\omega w_j^\dagger w_j ) + {i \over 2} {\rm Tr} \omega \hat{\gamma}_5 , 
\end{eqnarray}
where, as usual, the measure current is cancelled by the variation of the Jacobian and light partition function. The gauge variation of $Z_-$ is, naturally, the same as in (\ref{deltaZchiral}) (the gauge variation of $Z_+$ can be obtained similarly to (\ref{vartn1}) and be seen to combine, together with the Jacobian to yield (\ref{z9})).
 
 The determinant prefactor in $Z_-$ contributes both to the real and imaginary parts of the mirror polarization operator. It is clear from (\ref{vartn1}) above that the imaginary part of the polarization operator due to the determinant is exactly as required by anomaly cancellation.  Assuming   unitarity at long distances, one would  argue that Re$\Pi_{\mu\nu}$ should  receive a contribution from at least one massless charged chiral fermion (plus, possibly, a number of massless states in anomaly-free representations).   The goal of our simulation was to precisely find out the real part of the mirror polarization operator. 
 
 Let us now compare the numerical findings described in Section \ref{symmetryvsexplicit} with what can be inferred from the representation of the mirror partition function (\ref{Ztotal2}, \ref{Ztotal3}).  
 
 The real part of the mirror theory $\Pi_{\mu\nu}^-$ receives two contributions at $y=\infty$. The first is due to   the determinant prefactor. The contribution to Re$\Pi_{\mu\nu}^-$ of ${\rm det}_- (2 - D^1)$ can be seen\cite{Poppitz:2009gt} to be exactly that of   three massless propagating charged chiral fermions---the chiral components of the three 2d doubler modes. The determinant prefactor contribution to Re$\Pi_{\mu\nu}^-$ is, obviously, the same for any value of $y, h$.

 The only other  contribution to Re$\Pi_{\mu\nu}^-$ at $y=\infty$ arises 
from the nonlocal coupling in the Majorana Yukawa term in $S_{Yukawa}^\prime$  of (\ref{Ztotal1}). While it appears difficult to calculate analytically this contribution in the disordered-$\phi$ phase, 
our numerical results for the mirror polarization operator Re$\Pi_{\mu\nu}$  show  that there is {\it one} massless propagating chiral fermion. 
Thus  two of the three massless modes contributed by the prefactor are cancelled by the nonlocal contribution to Re$\Pi_{\mu\nu}$ from the Majorana term in $S_{Yukawa}$; it appears that this cancellation is exact for all values of $h>1$.
 
For vanishing Majorana coupling, $h=0$, on the other hand, both   numerical simulations (with the code of \cite{Giedt:2007qg} used, we  can only approach the $h \rightarrow 0$ limit) and  the representation (\ref{Ztotal2}) show that the real part of the mirror polarization operator at small momenta and infinite $y$ is that of three charged massless chiral fermions. This is not too surprising as the Majorana coupling goes away in the limit and at infinite $y$ the only dependence on the gauge field is in the determinant  prefactor in $Z_-$ of (\ref{Ztotal3}).
This indicates the  crucial role of the Majorana-type couplings (recall that they were motivated by the need to   break all mirror global symmetries) in facilitating the decoupling of the maximal possible number, as  allowed by anomaly matching, of charged mirror degrees of freedom.
 
  Another lesson we learned  is that probing the fermion spectrum with local fermion operators (including charged local fermion-scalar composites, as in\cite{Giedt:2007qg}) can miss massless degrees of freedom. The massless charged mirror fermions   were not seen in that study, perhaps because they are not  expressed in an obvious  local way through the original variables.
   The long-distance gauge-boson polarization operator of the mirror theory is a universal probe of the charged mirror spectrum and should  be the first quantity, along with susceptibilities probing  chiral symmetry breaking,  computed in any future studies of anomaly-free models.

\section{Outlook}
\label{summary}

We have seen that when GW chiral
fermions are used, the question of decoupling the mirror fermions 
from a vectorlike gauge theory is intimately intertwined with 
their contributions to anomalies. While this conclusion is not 
unexpected, these issues were usually not the central 
problem in earlier studies---largely because correlators of global 
chiral currents in the mirror theory could not be studied, as the  exact lattice
chiral symmetries were not known. This is where the GW formalism
exhibits its advantages: when it is used,
a precise formulation on the lattice of 't Hooft's anomaly 
matching condition can be found and verified. 
 
We have seen that models with 
mirror Yukawa interactions  formulated via exact lattice chirality 
are ``smart" enough that, consistent with anomaly matching, in the 
strong-coupling symmetric phase they lead to massless degrees of freedom as required by   anomaly matching, 
rather than to a nonunitary theory. In the  ``1-0" model, the minimal 
number of chiral fermions needed to cancel the light fermion anomaly remain 
massless in the strong-coupling symmetric phase when all mirror global symmetries are explicitly broken. 

An obvious  question is what these results imply for anomaly-free 
models. 

Take, for example,
 the ``3-4-5'' model, a two dimensional chiral $U(1)$ gauge theory with  $3_{-}$, $4_{-}$ and $5_{+}$ massless 
fermions, where the number indicates the $U(1)$ charge and $\pm$ the chirality (we stick to 2d anomaly-free examples, as they are the logical next step; it is not realistic to expect the mirror dynamics of, e.g., the 4d  SU(5) model of Section 
 \ref{chiralfromvectorlike} 
to be studied numerically soon.)

\begin{figure}
\begin{center}
\includegraphics[width=4in]{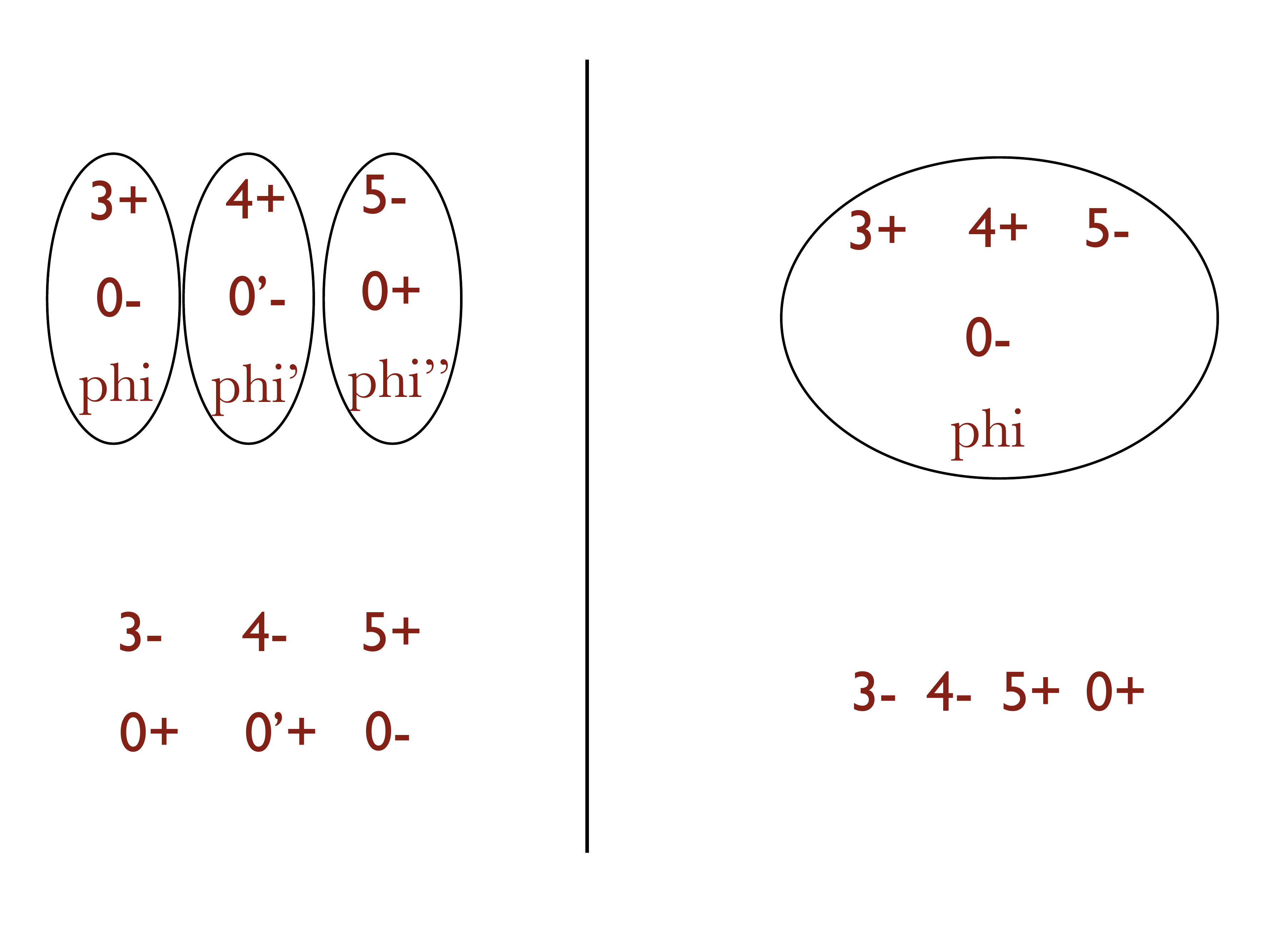}
\caption
{Two possible ways to  engineer  mirror-fermion dynamics in the ``345" model; the circles are meant to indicate that all interactions that break the mirror global symmetries  between the enclosed fields are included. On the left: three copies of the ``1-0" model leave three anomaly-free $U(1)$'s whose 't Hooft anomalies require the presence of at least three charged massless chiral mirror fermions. On the right: the only unbroken global mirror symmetry is the to-be-gauged $U(1)$; since the mirror spectrum is anomaly free, 't Hooft anomalies do not require presence of charged massless states. Conjecture (testable!): the extra interactions breaking all global symmetries give mass to all mirror massless states in the strong-coupling symmetric phase---``just like" including the Majorana coupling in the ``1-0" model lifted the massless vectorlike pair.
}
 \label{345story} 
\end{center}
\end{figure}

 Consider first implementing the decoupling or mirrors from 
a vectorlike theory with three Dirac fermions of charges  3, 4, 5 
by adding three uncharged mirror fermions and three unitary scalars, 
i.e.~taking three copies of the ``1-0" model and appropriately 
changing charges/chiralities (one could call the three copies the  
``3-0", ``4-0",  and ``5-0" models). This is indicated on the lhs of Fig.~\ref{345story}.
 It is clear that 
such an implementation of the ``3-4-5" model would lead to massless mirror 
states already in trivial gauge backgrounds,  despite two facts that might 
suggest otherwise:   that the mirror and light spectra are  anomaly-free 
and that, when the global $U(1)$ appropriate to yield the ``3-4-5"  model is 
gauged, both the anomaly-free and anomalous global symmetries of such a 
lattice implementation are as in the target continuum theory. The easiest 
way to see that massless modes will result at zero gauge background is to  
weakly  gauge the three separate global $U(1)$ symmetries present in the 
``3-0", ``4-0", and ``5-0" models and demand  consistency of the two-point current 
correlators, as we did for the ``1-0" model above (the polarization operators for
these   $U(1)$s are essentially a set of correlation
functions in the theory and we evaluate their divergences only at 
zero gauge-field background; thus all   conclusions of Section \ref{implementation} hold). 

The condition, already formulated in Section \ref{mirrorsymmetries}, but repeated here for completeness, that  the mirror interactions 
should obey in order to avoid massless states due to anomaly matching 
from an extra global symmetry is therefore to demand that when the 
gauge interactions 
are turned off, the mirror theory should have no global symmetries other 
than the global part of the gauge group---as elucidated in Section \ref{mirrorsymmetries}. With gauge interactions turned off, 
the implementation of the ``3-4-5" model of the previous paragraph has, instead, two 
extra global $U(1)$s, which act simultaneously on the light and mirror 
components and  impose further conditions on the mirror spectrum that 
imply  the existence of  massless modes.
This example  leads us to conjecture  that if the mirror interactions
couple the $3_{+}$, $4_{+}$, and $5_{-}$ mirrors, as schematically shown on the rhs of Fig.~\ref{345story},   by adding only one
scalar and a single neutral $0_{-}$ mirror fermion and including the 
most general gauge-invariant couplings breaking all mirror global 
chiral symmetries, 
 there wouldn't be any massless mirror states in 
the strong-coupling symmetric phase. 

At the moment, the strongest argument for the validity of our conjecture 
is that with all  mirror global symmetries explicitly broken,  
there is no reason we know of, such as anomaly cancellation of any 
symmetry, for massless mirrors to exist at strong coupling (note that strong coupling is a must: since  the gauge symmetry forbids mirror fermion mass terms, at weak mirror couplings and in an unbroken phase the mirror fermions will stay massless). 
The qualitative expectation at strong coupling is that the extra couplings on the rhs of Fig.~\ref{345story} (compared to the   lhs)
that break all extra global symmetries will give mass to the massless mirror states
that exist on the rhs (just like the Majorana coupling $h$ of the ``1-0" model  gave mass to the vectorlike pair which remained massless at $h=0$ in the strong symmetric phase, as  explained in Section \ref{domainwall}). Needless to say, understanding how this  mechanism works (or fails) in more detail would be very  interesting.

We end by noting that 
while achieving analytic understanding of the role of various couplings in the strong-coupling symmetric phase is quite desirable (but might be difficult), our conjecture about mirror-fermion decoupling can be  
tested via a  numerical ``experiment," 
similar to the one of\cite{Poppitz:2009gt}, but this time for an anomaly-free model. 
Since  no  dynamical gauge fields are needed at this stage, the study 
of anomaly-free mirror dynamics is  feasible  given appropriate 
computer resources and the techniques already developed in\cite{Giedt:2007qg,Poppitz:2007tu,Poppitz:2009gt} and reviewed here.

 \section*{Acknowledgments}

We thank   J. Giedt, S. Golkar, M. Golterman, R. Narayanan, A. Schwimmer, Y. Shamir,  J. Smit, and M. \" Unsal for discussions and comments.

\end{document}